\documentclass[twocolumn,english,aps,superscriptaddress,showpacs,prb]{revtex4}
\usepackage{array}
\usepackage{graphicx}
\usepackage{amssymb}
\usepackage{setspace}

\makeatletter
\usepackage{babel}

\begin{document}

\title{Hall of Mirrors Scattering from an Impurity in a Quantum Wire}

\author{J. Y. Vaishnav}

\email{vaishnav@physics.harvard.edu}

\affiliation{Harvard University Department of Physics, Cambridge MA 02138}

\author{A. Itsara}

\affiliation{Harvard College, Cambridge MA 02138}

\author{E. J. Heller}

\affiliation{Harvard University Department of Physics, Cambridge MA 02138}

\affiliation{Harvard University Department of Chemistry and Chemical Biology,
Cambridge MA 02138}

\begin{abstract}
This paper develops a scattering theory to examine how point impurities
affect transport through quantum wires. While some of our new results
apply specifically to hard-walled wires, others--for example, an effective
optical theorem for two-dimensional waveguides--are more general.
We apply the method of images to the hard-walled guide, explicitly
showing how scattering from an impurity affects the wire's conductance.
We express the effective cross section of a confined scatterer entirely
in terms of the empty waveguide's Green's function, suggesting a way
in which to use semiclassical methods to understand transport properties
of smooth wires. In addition to predicting some new phenomena, our
approach provides a simple physical picture for previously observed
effects such as conductance dips and confinement-induced resonances.
\end{abstract}

\pacs{72.10.Fk, 03.65.Nk, 73.23.-b}

\maketitle

\section{Introduction\label{sec:Introduction}}

Elastic scattering from a point defect in a hard-walled, multimode
quantum waveguide is a problem which has been the subject of experimental\cite{impurityExperiment1,impurityExperiment2}
and theoretical\cite{ImpurityTheoryReichl2,ImpurityTheory2004,ImpurityTheoryAJP,ImpurityTheoryReichl,ImpurityTheory1,ImpurityTheoryBagwell,ImpurityImages,Datta}
inquiry in both condensed matter and atomic physics. In this paper,
we revisit the problem from an unconventional point of view, reducing
it to the scattering of a single effective wavefunction off an array
of images. Our approach enables us to easily understand complex transport
phenomena which have already been observed (such as conductance dips),
as well as predict other new ones.

The problem of scattering from pointlike impurities in waveguides
is a general one, and thus has physical applications in several fields.
Impurity scattering in quantum wires, and the resulting disorder effects
on electron transport, have long been of interest in mesoscopic physics.
Specifically, the problem arises in the context of two-dimensional
electron gases (2DEGs) confined by quantum wire potentials; the impurity
can represent a defect in the wire. Such quantum waveguides can be
fabricated at low-temperature ${\rm Ga}_{1-x}{\rm Al}_{x}{\rm As}$
interfaces, and are of interest due to their potential role in high-frequency
and quantum devices. As carbon nanotubes can behave as few-mode quantum
waveguides,\cite{ParkNanotube} our results are relevant to transport
in nanotubes with adsorbed impurities--of interest in the study of
biosensors. The problem of scattering in confined geometries arises
in atom waveguides as well, with potential implications for quantum
computing and atom interferometry. In both cases, atoms must maintain
coherence as they pass through the waveguide, and the goal is thus
to minimize phenomena such as collisional phase shifts.\cite{Olshanii}
Finally, the wave phenomena arising in quantum waveguides are in direct
mathematical correspondence with a number of wave phenomena in other
systems: Water waves resonantly trapped between an array of cylinders,\cite{Water}
large antenna arrays,\cite{Antennas} and sound waves directed by
``Bessel'' line arrays in acoustics. \cite{Acoustics}

\begin{figure}[htb]
\begin{center}\includegraphics[%
  width=0.8\columnwidth,
  keepaspectratio]{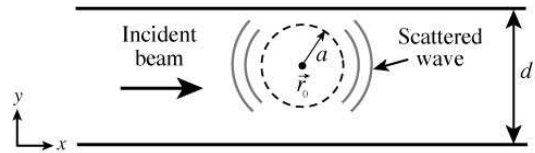}\end{center}
\caption{Schematic of the quantum wire, with a point impurity of scattering
length $a$ at $\vec{r}_{0}=(x_{0},y_{0}).$ We assume infinite leads,
and hard walls at $y=0$ and $y=d$. \label{fig:schematic}}
\end{figure}

Motivated by the above applications, we examine low-energy scattering
of noninteracting particles from impurities in an infinite two-dimensional
quantum wire, with hard walls at $y=0$ and $y=d.$ The wire (Fig.
\ref{fig:schematic}) contains a point impurity, which we model as
an $s$ wave scatterer of effective radius $a$. %
Our method differs from previous treatments, such as the renormalized
$t$ matrix method,\cite{AdamThesis,Olshanii,Olshanii2} in that we
use an unconventional approach which combines the following three
ideas: 

\begin{enumerate}
\item The method of images.
\item The realization that a confined scatterer, like a free space scatterer,
is a rank-one target. We can, therefore, combine the $N$ degenerate
transverse modes into a basis in which only a single wavefunction
scatters.
\item That the effect of scattering from the waveguide walls is to renormalize
this single scattering wavefunction, so that it takes on a new effective
value at the scatterer location.
\end{enumerate}
Using the method of images allows us to explicitly understand how
each reflection from the waveguide walls affects the conductance.
The fact that the confined scatterer is a rank one target enables
us to express the transport properties of the wire entirely in terms
of the single scattering wavefunction. Finally, the idea of renormalization
allows us to fully explain the transport properties of the wire in
terms of reflections of this single wavefunction from the waveguide
walls.

In Section \ref{sec:freeSpace}, we review free space multiple scattering,
anticipating phenomena which will appear in the wire. Section \ref{sec:Wire}
lays out a similar formalism for scattering in the waveguide. In Section
\ref{sec:Results-and-Discussion:}, we present physical phenomena
which we observed in our model, which we interpret in Section \ref{sec:Physical}
in terms of interference effects. We relegate mathematical details
to Appendices \ref{sec:diffraction}-\ref{sec:kummer}.

\section{Preliminaries: Review of Free Space Scattering\label{sec:freeSpace}}

In Section \ref{sec:Wire}, we show that the Lippmann-Schwinger equation
for scattering from a confined impurity resembles the free space Lippmann-Schwinger
equation, with the exception that in the wire, a renormalized incoming
wavefunction replaces the true one. Surprisingly, even for a single
confined scatterer, this renormalized wavefunction includes free space
\emph{multiple} scattering effects. We thus begin by briefly reviewing
free space scattering, emphasizing the three phenomena which will
reappear in the Lippmann-Schwinger equation for a single scatterer
in the waveguide.

Consider $N$ pointlike impurities in two-dimensional free space,
located at $\left\{ \vec{r}_{1},\dots,\vec{r}_{N}\right\} .$ Let
$V(\left|\vec{r}-\vec{r}_{i}\right|)$ be the potential at $\vec{r}$
due to the $i^{{\rm th}}$ scatterer. We seek scattering solutions
$\psi(\vec{r})$ to the time-independent Schrodinger equation\begin{equation}
\left[-\frac{\hbar^{2}}{2m}\vec{\nabla}^{2}+\sum_{i=1}^{N}V(\left|\vec{r}-\vec{r}_{i}\right|)-E\right]\psi(\vec{r})=0,\label{eq:schrodinger}\end{equation}
 under the assumption of low-energy, $s$-wave scattering. Except
as noted, we henceforth use atomic units, where $\hbar=m=c=1.$ Applying
the $t$ matrix formalism,\cite{tmatrix} we characterize a single
scatterer at $\vec{r}_{i}$ by its $t$ matrix,\begin{equation}
t=s(k)\left|\vec{r}_{i}\rangle\langle\vec{r}_{i}\right|\label{eq:tmatrix}\end{equation}
where $s(k)$ is a function of the wavenumber, $k=\sqrt{2mE}/\hbar.$
We choose the functional form of $s(k)$ to simulate the scatterer
of interest, under the constraint that $s(k)$ must satisfy the optical
theorem

\begin{equation}
{\rm -2Im}s(k)=\left|s(k)\right|^{2}.\label{eq:optical1}\end{equation}
In this paper, we choose\begin{equation}
s(k)=-2i\frac{\hbar^{2}}{m}\frac{J_{0}(ka)}{H_{0}^{(1)}(ka)},\label{eq:tmatrix2}\end{equation}
which represents a hard disk of scattering length $a$.

\subsection{Single Scatterer\label{sub:Single-Scatterer}}

Consider a single scatterer at $\vec{r}=\vec{r}_{0}.$ In our $t$
matrix representation, an incident wave $\phi(\vec{r})$ scatters
into a wavefunction $\psi(\vec{r})$ according to the Lippmann-Schwinger
equation\begin{equation}
\psi(\vec{r})=\phi(\vec{r})+s(k)\phi(\vec{r}_{0})G_{0}(\vec{r},\vec{r}_{0};k)\label{eq:LippmannSchwingerSingle}\end{equation}
where

\begin{equation}
G_{0}(\vec{r},\vec{r}_{0};k)=\frac{2m}{\hbar^{2}}\left[-\frac{i}{4}H_{0}^{(1)}(k|\vec{r}-\vec{r}_{0}|)\right]\label{eq:freeGreens}\end{equation}
is the 2D free space advanced Green's function satisfying

\begin{equation}
\left(\vec{\nabla}^{2}+k^{2}\right)G_{0}(\vec{r},\vec{r}_{0};k)=\frac{2m}{\hbar^{2}}\delta^{(2)}(\vec{r}-\vec{r}_{0}).\label{eq:greens}\end{equation}
We shall henceforth assume we are using the advanced Green's function
and omit the superscript on the Hankel function; we shall also omit
the implicit $k$ dependence in the Green's function.

\subsubsection*{Single Scattering Wavefunction\label{sub:Single-Scattering-Wavefunction2}}

Any incoming wave $\phi(\vec{r})$ must be a solution to the free
space Schrodinger equation. We can therefore express $\phi(\vec{r})$
in the basis of ``cylinder harmonics,'' \begin{equation}
\phi(\vec{r})=\sum_{m=0}^{\infty}c_{m}J_{m}(k|\vec{r}-\vec{r}_{0}|)\times\frac{e^{im(\phi-\phi_{0})}}{\sqrt{2\pi}}\label{eq:freespaceswf}\end{equation}
where $m=0,1,2,\dots$ correspond to $s$,$p,$$d,\dots$ waves. An
$s$ wave scatterer in free space is a rank one perturbation: Of all
the terms in (\ref{eq:freespaceswf}), only the $m=0$ term is nonzero
at the scatterer. From (\ref{eq:LippmannSchwingerSingle}), then, only
the $s$ wave scatters, acquiring a phase shift; remaining higher
partial waves pass through the scatterer unperturbed. While trivial
in free space, these facts will reappear more subtly in the wire.

\subsection{Multiple Scatterers\label{sub:freeSpaceMultiple}}

The Lippmann-Schwinger equation describing scattering from a collection
of $N$ identical point scatterers at $\left\{ \vec{r}_{1},\dots,\vec{r}_{N}\right\} $
is\begin{equation}
\psi(\vec{r})=\phi(\vec{r})+s(k)\sum_{i=1}^{N}\psi_{i}(\vec{r}_{i})G_{0}(\vec{r},\vec{r}_{i})\label{eq:lippmannSchwingerMultiple1}\end{equation}
 in which we express the $\psi_{i}(\vec{r}_{i})$ recursively as\begin{equation}
\psi_{i}(\vec{r}_{i})=\phi(\vec{r}_{i})+s(k)\mathop{\sum_{j=1}^{N}}_{j\neq i}\psi_{j}(\vec{r}_{j})G_{0}(\vec{r}_{i},\vec{r}_{j}).\label{eq:lippmannSchwingerMultiple2}\end{equation}
where we have followed Foldy's method.\cite{Foldy} Comparing (\ref{eq:lippmannSchwingerMultiple1}-\ref{eq:lippmannSchwingerMultiple2})
with the single-scatterer version (\ref{eq:LippmannSchwingerSingle}),
the $\psi_{i}(\vec{r}_{i})$ are the $effective$ incoming wavefunctions
at each scatterer: $\psi_{i}(\vec{r}_{i})$ is the amplitude incident
on the $i^{{\rm th}}$ scatterer after scattering from each of the
other scatterers. A crucial point is that $\psi_{i}(\vec{r}_{i})$
excludes the singular self-interaction of the $i^{\textrm{th}}$ scatterer. 

Defining $\vec{\phi}_{i}\equiv\phi(\vec{r}_{i}),$ $\vec{\psi}_{i}\equiv\psi(\vec{r}_{i}),$
inverting (\ref{eq:lippmannSchwingerMultiple1}-\ref{eq:lippmannSchwingerMultiple2})
yields\begin{equation}
\vec{\psi}=(\mathbf{1}-s\mathbf{G})^{-1}\vec{\phi}\label{eq:lippmannSchwingerMatrix}\end{equation}
where\begin{equation}
G_{ij}\equiv\left\{ \begin{array}{cc}
G_{0}(\vec{r_{i}},\vec{r}_{j}) & i\neq j\\
0 & i=j\end{array}\right.\label{eq:greensmatrixeq}\end{equation}
excludes the singular term. An alternate expression of (\ref{eq:lippmannSchwingerMatrix})
is as the Born series\begin{equation}
\vec{\psi}=\left[\mathbf{1}+s\mathbf{G}+(s\mathbf{G})^{2}+(s\mathbf{G})^{3}+\dots\right]\vec{\phi}.\label{eq:expansion}\end{equation}
The effective wavefunctions thus have a simple interpretation in terms
of interfering paths: The terms in square brackets describe amplitude
incident at $\vec{r}_{i}$ after interactions with zero, one, two,
or three other scatterers respectively. The series continues infinitely.

Anticipating phenomena that will reappear in the wire, we highlight
the following points of this section:

\begin{enumerate}
\item The effect of multiple scattering is to create a new effective incoming
wavefunction at each scatterer.\label{enu:newWf}
\item The effective wavefunction at a particular scatterer is a sum of waves
scattered from all the \emph{other} scatterers, and excludes the (singular)
self-interaction of the scatterer.\label{enu:renorm}
\item An $s$ wave scatterer in free space is a rank one target.\label{enu:specialWf}
\end{enumerate}

\section{Scatterer in a wire\label{sec:Wire}}

In this section, we examine how confinement in a wire affects the
transport properties of a scatterer. Applying the method of images,
we derive a form of the empty wire Green's function. Using this form
of the Green's function, we show that the sole effect of confinement
is to renormalize the incoming wavefunction. The image representation
allows us to very simply calculate the renormalization coefficient.
We define an effective cross section for a confined scatterer, and
relate it to the conductance of the wire-impurity system. We derive
an effective optical theorem in the wire. Finally, applying this formalism,
we investigate the behavior of the cross section and conductance as
functions of various parameters.

\subsection{Green's Function via the Method of Images \label{sec:greensFunction}}

In order to write down the Lippmann-Schwinger equation for an impurity
in the wire, we require the Green's function $G_{w}(\vec{r},\vec{r}_{0})$
of the empty wire. The usual spectral form of the empty wire Green's
function is\begin{equation}
G_{w}(\vec{r},\vec{r}_{0})=-i\sum_{m=1}^{\infty}\frac{1}{k_{x}^{(m)}}\chi_{m}(y)\chi_{m}(y_{0})e^{ik_{x}^{(m)}|x-x_{0}|}\label{eq:dattagreenswire}\end{equation}
where the $\chi_{m}(y)$ are the transverse modes for the particular
waveguide (see e.g. Datta\cite{Datta}). In our hard wire, the modes
have the form \begin{equation}
\begin{array}{ccc}
\chi_{m}(y) & = & \left\{ \begin{array}{ll}
\sqrt{\frac{2}{d}}\sin\left(\frac{m\pi y}{d}\right) & 0<y<d\\
0 & y<0,\, y>d.\end{array}\right.\end{array}\label{eq:wireModes}\end{equation}
We note that the spectral Green's function (\ref{eq:dattagreenswire})
is a sum over evanescent modes as well as propagating ones.

Although we could, in principle, proceed using the spectral form (\ref{eq:dattagreenswire})
of the Green's function, the spectral Green's function provides little
physical insight into the precise mechanism of the scattering. We
therefore take an alternative approach, and use the method of images
to derive an alternate, equivalent form of (\ref{eq:dattagreenswire})
for the hard wire.

Although the problem of an impurity in a hard wire has been the subject of numerous theoretical studies,\cite{ImpurityTheory1,ImpurityTheory2004,ImpurityTheoryAJP,ImpurityTheoryBagwell,ImpurityTheoryReichl,ImpurityTheoryReichl2,ImpurityImages}
the authors are aware of only one work\cite{ImpurityImages} which applies 
the method of images:  Ref. 3 treats the related problem of a point scatterer in a 3D waveguide of rectangular cross section, which the authors reduce to the problem of scattering from an impurity in a finite, 2D box. The method of images has the significant advantage of making multiple scattering effects
explicit.  We shall thus use the approach of Ref. 3, rather than the more common spectral methods.  Because our problem of an infinite waveguide is somewhat simpler than the finite box treated in Ref. 9, the role of multiple scattering is more transparent. We are therefore able to explore in detail the considerable effect of multiple scattering on the physical properties of the wire, which, although briefly mentioned, is not examined in Ref. 3.

The empty wire Green's function $G_{w}(\vec{r},\vec{r}_{0})$ satisfies
Green's equation inside the wire, and is zero on the wire walls:

\begin{eqnarray}
(\vec{\nabla}^{2}+k^{2})G_{w}(\vec{r},\vec{r}_{0}) & = & 2\delta(\vec{r}-\vec{r}_{0})\label{eq:greenWire}\\
G_{w}(x,0) & = & G_{w}(x,d)=0\label{eq:wireBC}\end{eqnarray}
The confined Green's function differs from the free space version
due to the necessity of including reflections off the walls when describing
the response to a point excitation. Note the factor of two in (\ref{eq:greenWire}),
obtained by casting (\ref{eq:greens}) in atomic units.

\begin{figure}[htb]
\begin{center}\includegraphics[%
  width=0.5\columnwidth]{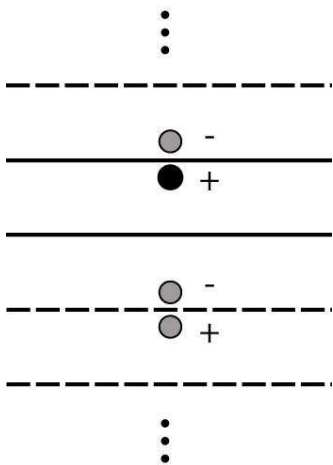}\end{center}
\caption{The method of images allows us to reduce the problem of a single
scatterer in a wire to that of a periodic array of scatterers in free
space. The solid lines and black scatterer are the actual wire and
point source. The gray point sources, and the dashed lines, are images.
The pluses and minuses refer to the sign of the free-space Green's
function term that the particular point source contributes. The alternating
signs of these contributions are due to the Dirichlet boundary conditions;
the wavefunction must cancel on the wire walls. \label{fig:offCenter}}
\end{figure}

As in electrostatics, we can linearly combine free space Green's functions
(\ref{eq:freeGreens}) with point sources in different places, so
that the linear combination satisfies (\ref{eq:wireBC}). As long
as only one of the point sources is inside the wire $(0<y<d)$, our
sum will satisfy (\ref{eq:greenWire}) inside the wire. Outside the
wire, we discard the solution. If the point source is located at $\vec{r}=(x_{0},y_{0})$,
we use the configuration of image point sources in Fig. \ref{fig:offCenter}.
We reflect the scatterer, creating a series of images such that the
position of the $n^{\mathrm{th}}$ image is $\vec{r}_{n}=x_{0}\hat{x}+[(-1)^{n}y_{0}+2nd]\hat{y}.$

This image configuration yields an empty wire Green's function of
the form
\begin{equation}
G_{w}(\vec{r},\vec{r}_{0})=\sum_{n=-\infty}^{\infty}(-1)^{n}G_{0}(\vec{r,}\vec{r}_{n})\label{eq:wireGreens}\end{equation}
 The Green's function (\ref{eq:wireGreens}) satisfies (\ref{eq:greenWire})
inside the wire, and (\ref{eq:wireBC}) on the wire boundaries. Outside
the wire we implicitly set the Green's function to zero. The Green's
function satisfies the Dirichlet boundary conditions because, as we
take infinitely many images, the alternating sum of Hankel functions
converges to zero on the walls. This series converges extremely slowly,
precluding its use for numerical purposes (see the discussion in Morse
and Feshbach\cite{MorseFeshbach}).

\subsection{Renormalizing the Incoming Wavefunction\label{sub:Renormalizing-the-Incoming}}

The Lippmann-Schwinger equation requires knowledge of both the Green's
function and the incoming wavefunction. In Section \ref{sub:Single-Scatterer},
we calculated the Green's function for the wire. In this section,
we discuss the incident wavefunction. We show that the effects of
confinement resemble free space multiple scattering: Confinement renormalizes
the effective wavefunction at the scatterer from its true value. In
the image formalism, the multiple scattering arises due to scattering
off images.

Applying the free space multiple scattering formalism (\ref{eq:lippmannSchwingerMultiple1},\ref{eq:lippmannSchwingerMultiple2})
to the array of images in Fig. \ref{fig:offCenter}, we find that
the Lippmann-Schwinger equation for the wire is\begin{eqnarray}
\psi(\vec{r}) & = & \phi(\vec{r})+s\sum_{i=-\infty}^{\infty}G_{0}(\vec{r},\vec{r}_{i})\psi_{i}(\vec{r}_{i})\label{eq:ms}\end{eqnarray}
where\begin{equation}
\psi_{i}(\vec{r}_{i})=\phi(\vec{r}_{i})+s\mathop{\sum_{j=-\infty}^{\infty}}_{j\neq i}G_{0}(\vec{r}_{i},\vec{r}_{j})\psi_{j}(\vec{r}_{j}).\label{eq:wavelets1}\end{equation}
Recall from our discussion of free space multiple scattering the physical
meaning of this recursive Lippmann-Schwinger equation: The $\psi_{i}(\vec{r}_{i})$
defined in (\ref{eq:wavelets1}) are the effective incoming wavefunctions
at the $i^{\textrm{th}}$ scatterer in the array. In a general multiple
scattering problem, we would have to stop here and solve numerically.
The image array is, however, a special case: Not only is it periodic,
but the boundary conditions at the wall impose the antisymmetry\begin{eqnarray}
\phi(\vec{r}_{i}) & = & (-1)^{i}\phi(\vec{r}_{0})\nonumber \\
\psi_{i}(\vec{r}_{i}) & = & (-1)^{i}\psi_{0}(\vec{r}_{0}).\label{eq:as}\end{eqnarray}
Any wavefunction, incident or scatterered, must be zero on the walls.
Thus, inside the wire, in the $y$ direction, the wavefunction is
a superposition of modes (\ref{eq:wireModes}). Replacing the wire
with an image array requires that we extend these modes outside the
wire, into the region $y<0,\, y>d.$ Keeping the $\chi_{m}(y)$ continuous
at the walls, we can extend them so that the wavefunction is either
symmetric or antisymmetric across the wire walls; either set forms
a basis. However, we know that in the absence of a wall $\chi_{m}^{\prime}(y)$
must be continuous also, as we have eliminated the hard walls. The
symmetric extension does not satisfy this condition. We therefore
discard the symmetric extension, leaving (\ref{eq:as}) as the proper
boundary condition.

Combining (\ref{eq:wavelets1}) and (\ref{eq:as}), the equation for
the wavelets thus becomes\begin{eqnarray}
\psi_{i}(\vec{r_{i}}) & = & (-1)^{i}\left[\phi(\vec{r}_{0})+s\psi_{0}(\vec{r}_{0})G_{r}\right]\label{eq:wavelets}\end{eqnarray}
where we have defined the renormalization sum\begin{eqnarray}
G_{r} & \equiv & \mathop{\sum_{i=-\infty}^{\infty}}_{i\neq0}(-1)^{i}G_{0}(\vec{r_{i}},\vec{r}_{0})\label{eq:lambda}\end{eqnarray}
noting that\begin{equation}
G_{r}=(-1)^{j}\mathop{\sum_{i=-\infty}^{\infty}}_{i\neq j}(-1)^{i}G_{0}(\vec{r_{i}},\vec{r}_{j}).\label{eq:signstrength}\end{equation}

Taking $i=0$ in (\ref{eq:wavelets}), we find \begin{equation}
\psi_{0}(\vec{r}_{0})=\frac{\phi(\vec{r}_{0})}{1-sG_{r}}\label{eq:zerowavelet}\end{equation}
Substituting (\ref{eq:as},\ref{eq:zerowavelet}) into (\ref{eq:ms})
yields the Lippmann-Schwinger equation for scattering from an impurity
in a wire:\begin{eqnarray}
\psi(\vec{r}) & = & \phi(\vec{r})+s\left[\frac{\phi(\vec{r}_{0})}{1-sG_{r}}\right]\sum_{i=-\infty}^{\infty}(-1)^{i}G_{0}(\vec{r},\vec{r}_{i})\label{eq:msWf}\\
 & = & \phi(\vec{r})+s\left[\frac{\phi(\vec{r}_{0})}{1-sG_{r}}\right]G_{w}(\vec{r},\vec{r}_{0}).\label{eq:msWf2}\end{eqnarray}
Comparing (\ref{eq:msWf2}) to the free space Lippmann-Schwinger equation
(\ref{eq:LippmannSchwingerSingle}) for a single scatterer, we see
that the wire simply renormalizes the incoming wavefunction at the
scatterer to have a new effective value \begin{equation}
\tilde{\phi}(\vec{r}_{0})=\frac{\phi(\vec{r}_{0})}{1-sG_{r}}\label{eq:renormalizedSwfGr}\end{equation}
and so our final Lippmann-Schwinger equation is\begin{equation}
\psi(\vec{r})=\phi(\vec{r})+s\tilde{\phi}(\vec{r}_{0})G_{w}(\vec{r},\vec{r}_{0}).\label{eq:lippmannschwingerwireagain}\end{equation}
For convenience, we define a renormalization factor\begin{equation}
\mathcal{R}\equiv\frac{\phi(\vec{r}_{0})}{\tilde{\phi}(\vec{r}_{0})}\label{eq:renormalizationfactorR}\end{equation}
which is the ratio between the original and renormalized incoming
wavefunctions at the scatterer.

\begin{figure}[htb]
\begin{center}\includegraphics[%
  width=0.7\columnwidth,
  keepaspectratio]{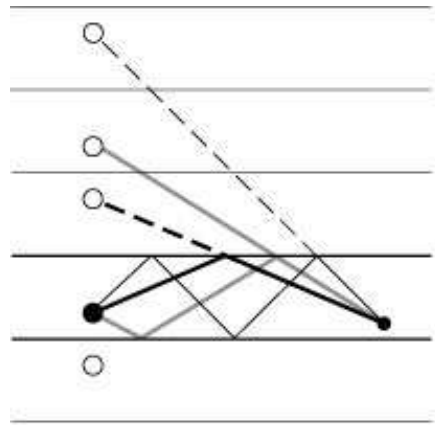}\end{center}

\caption{Amplitude scattering from the $n^{th}$ image corresponds to amplitude
reaching a point after $n$ bounces from the wall, each of which reverses
the sign of the wavefunction (phrased semiclassically, reflections
introduce a Maslov index of -1). Resonances occur when the interference
of all paths is maximally constructive at the scatterer itself.\label{fig:Born}}
\end{figure}

In order to cast the renormalized incoming wavefunction in terms of
interfering paths, we expand perturbatively, as we did in (\ref{eq:expansion})
for free space. For shorthand, define\begin{equation}
G_{i\rightarrow j}\equiv(-1)^{i+j}G(\vec{r}_{i},\vec{r}_{j}).\label{eq:propagateitoj}\end{equation}
We find\begin{eqnarray}
\tilde{\phi}(\vec{r}_{0}) & = & (1+sG_{r}+s^{2}G_{r}^{2}+\dots)\phi(\vec{r}_{0})\nonumber \\
 & = & \left[1+\mathop{\sum_{i=-\infty}^{\infty}}_{i\neq0}sG_{i\rightarrow0}\right.\nonumber \\
 &  & +\left.\mathop{\sum_{i=-\infty}^{\infty}}_{i\neq0}\mathop{\sum_{j=-\infty}^{\infty}}_{j\neq0}sG_{i\rightarrow0}\times sG_{j\rightarrow0}+\dots\right]\phi(\vec{r}_{0})\nonumber \\
 & = & \left[1+\mathop{\sum_{i=-\infty}^{\infty}}_{i\neq0}sG_{i\rightarrow0}\right.\label{eq:bouncesFeynman}\\
 &  & \left.+\mathop{\sum_{i=-\infty}^{\infty}}_{i\neq j}\mathop{\sum_{j=-\infty}^{\infty}}_{j\neq0}sG_{i\rightarrow j}\times sG_{j\rightarrow0}\right]\phi(\vec{r}_{0})\nonumber \end{eqnarray}
 where we have used (\ref{eq:signstrength}). The $n^{{\rm th}}$
term of the sum (\ref{eq:bouncesFeynman}) describes propagation to
$\vec{r}_{0}$ after scattering from any $n-1$ images, beginning
with the first term, which describes free propagation to $\vec{r}_{0}$.
Alternately, the $n^{{\rm th}}$ term corresponds to $n$ scattering
events combined with any number of reflections off the wall, each
of which introduces an additional phase of -1. Fig. \ref{fig:Born}
illustrates the correspondence between scattering from an image and
reflecting from the wall.

Renormalization of the effective wavefunction is simply the manifestation
of the identical phenomenon in free space (see item \ref{enu:newWf}
in the list of Section \ref{sub:freeSpaceMultiple}). From a semiclassical
point of view, the incoming wave, given infinite time to spread, reflects
from each possible combination of scatterers before returning to the
source. Note that $G_{r}$ is simply the wire's Green's function evaluated
at the scatterer, with the singular self-interaction of the source
removed,\begin{equation}
G_{r}=\lim_{\vec{r}\rightarrow\vec{r}_{0}}\left[G_{w}(\vec{r},\vec{r}_{0})-G_{0}(\vec{r},\vec{r}_{0})\right].\label{eq:gr}\end{equation}
This idea, as expressed in (\ref{eq:gr}), is the essence of  ``renormalized
$t$ matrix theory.''\cite{AdamThesis,Olshanii,Olshanii2} The theory,
as applied to our case, is equivalent to the idea that in free space
multiple scattering, the effective wavefunction at a scatterer excludes
the singular self-interaction (see item \ref{enu:renorm} of Section
\ref{sub:freeSpaceMultiple}). We note that although we have derived
(\ref{eq:gr}) via images only for the special case of a hard walled
guide, the result (usually derived perturbatively) is in fact true
for an arbitrary guide or confining potential.\cite{AdamThesis}

\subsection{Method of Images vs. Spectral Formulation}

We have presented both the image series (\ref{eq:wireGreens}) and
the spectral series (\ref{eq:dattagreenswire}) for the Green's function.
Because the Green's function of a system is unique, these two forms
must be equivalent. However, the physical relationship between the
method of images and the spectral form is far from obvious. The two
forms of the Green's function highlight different physical phenomena.
For example, reflections from the walls, which appear immediately
in (\ref{eq:wireGreens}), are far less obvious in the spectral form
(\ref{eq:dattagreenswire}). Contrastingly, the role of evanescent
channels in the scattering, while straightforward in the spectral
Green's function (\ref{eq:dattagreenswire}), is less transparent
in the image expansion (\ref{eq:wireGreens}) of the same Green's
function (although even the image expansion suggests that evanescent
channels will be present in some form, because Hankel functions are
singular). In order to understand the connection between the image
and spectral forms of the Green's function, we show their equivalence
mathematically in Appendix \ref{sec:diffraction}.

In Appendix A, we use an integral form of the Hankel function to show
the equivalence of (\ref{eq:wireGreens}) and (\ref{eq:dattagreenswire}).
The physical connection between the image and spectral formalisms
is diffraction. The images in Fig. \ref{fig:offCenter} form a periodic
array, which is effectively a diffraction grating. We note one small
difference from a typical diffraction situation: Due to the wire boundary
conditions, the incident modes are superpositions of two plane waves
each, rather than a single incident plane wave. This difference is
trivial, due to the linearity of Schrodinger's equation. When such
modes are incident on the effective lattice of images, they diffract:
Plane waves (or superpositions of plane waves) striking the grating
scatter, at large distances, into sums of plane waves at the Bragg
angles. The diffracted spectral orders are evident as the quantized
wavenumbers appearing in the spectral form of the Green's function.
As the Green's function for each image scatterer is singular, evanescent
waves appear in the Fresnel regime, near the lattice. We examine these
statements rigorously in Appendix \ref{sec:diffraction}.

\subsubsection*{Numerics}

While the method of images led quickly to the central results (\ref{eq:msWf2},\ref{eq:gr}),
these results expressed in terms of image sums converge so slowly
as to be impractical for numerics. Although the spectral form (\ref{eq:dattagreenswire})
converges more rapidly than the image series (\ref{eq:wireGreens}),
its convergence is still not uniform, as the evanescent modes include
a logarithmic singularity.

In Appendix \ref{sec:kummer}, we use Kummer's method to accelerate
the convergence of the Green's function, casting it in the form (\ref{eq:limitform})
suitable for numerical work. A side benefit of applying Kummer's method
is that we obtain a more rapidly converging expression for $G_{r}$:
In Appendix \ref{sec:kummer}, we show that the series (\ref{eq:lambda})
for $G_{r}$ (which is a Schl\"{o}milch series) resums to\begin{eqnarray}
G_{r} & = & \sum_{m=1}^{\infty}\left(\frac{1}{ik_{x}^{(m)}}+\frac{d}{m\pi}\right)\chi_{m}^{2}(y_{0})\nonumber \\
 &  & -\frac{1}{\pi}\ln\left[\frac{kd}{\pi}\sin\left(\frac{\pi y_{0}}{d}\right)\right]+\frac{i}{2}-\frac{\gamma}{\pi}\label{eq:schlomilchgreens}\end{eqnarray}
where $\gamma$ is the Euler-Mascheroni constant. We note that this
expression is considerably more complex than the expression (\ref{eq:lambda})
derived via the method of images. In the image formalism, the expression
for the renormalization constant $G_{r}$ has the intuitive form (\ref{eq:lambda}),
because removing the contribution of the source simply involves excluding
the scatterer itself, while retaining the images. In the spectral
formalism, because the singularity is not explicit, its removal is
considerably less transparent.

\subsection{Scattering Phenomena in Quantum Wires\label{sub:Scattering-in-Wire}}

In this subsection we define an effective cross section and optical
theorem for the confined impurity. As we have shifted from the image
formalism (which applies only to the hard wire) to the spectral one,
where (\ref{eq:dattagreenswire}) applies generally, the results presented
in Section \ref{sub:Scattering-in-Wire} apply to arbitrary waveguides,
not only hard wires.

\subsubsection{The S Matrix}

Since the wire is infinite, for calculating transmission/reflection
at infinity, we need consider only the open channels. Without loss
of generality (due to translational invariance of the wire in the
$x$ direction), suppose the scatterer is at $x_{0}=0.$ With incident
mode $n$, normalized to unit flux, the wavefunction at large distances
from the scatterer is\begin{eqnarray}
\psi(x,y;0,y_{0}) & = & \frac{e^{\pm ik_{x}^{(n)}x}}{\sqrt{k_{x}^{(n)}}}\chi_{n}(y)-\sum_{m=1}^{\infty}\left[\frac{i\mathcal{R}s}{\sqrt{k_{x}^{(n)}k_{x}^{(m)}}}\right.\nonumber \\
 &  & \left.\times\chi_{n}(y_{0})\chi_{m}(y_{0})\frac{e^{ik_{x}^{(m)}|x|}}{\sqrt{k_{x}^{(m)}}}\chi_{m}(y)\right]\label{eq:larger}\end{eqnarray}
from which we can read off a scattering matrix\begin{equation}
\textbf{S}=\left(\begin{array}{cc}
\mathbf{R} & \mathbf{T}'\\
\mathbf{T} & \mathbf{R}'\end{array}\right)=\left(\begin{array}{cc}
\mathbf{R} & \mathbf{I-R}\\
\mathbf{I-R} & \mathbf{R}\end{array}\right)\end{equation}
where the reflection and transmission coefficients are\begin{eqnarray}
R_{mn} & = & R'_{mn}=-\frac{i\mathcal{R}s}{\sqrt{k_{x}^{(n)}k_{x}^{(m)}}}\chi_{n}(y_{0})\chi_{m}(y_{0})\label{eq:rmatrix}\\
T_{mn} & = & T'_{mn}=\delta_{mn}-R_{mn}\label{eq:trmatrix}\end{eqnarray}
One can show algebraically (or verify numerically) that ${\rm rank(\mathbf{R})={\rm rank}(\mathbf{T-I)}=1.}$

\subsubsection{Defining a Cross Section in the Wire \label{sub:Effective-Cross-Section}}

In this section, we define an effective scattering cross section for
the wire. The usual three-dimensional free space differential scattering
cross section is\begin{equation}
\frac{d\sigma}{d\Omega}=\frac{dN(\Omega)}{N_{in}d\Omega}\end{equation}
where $dN(\Omega)$ is the number of particles per area scattered
into the solid angle $d\Omega$, and $N_{in}$ is the number of incident
particles per unit area. Expressed in terms of probability currents,
the above relation becomes\begin{equation}
\frac{d\sigma}{d\Omega}=\frac{\int j_{r}r^{2}d\Omega}{j_{in}}\end{equation}
where $\vec{j}={\rm Im}(\Psi^{*}\nabla\Psi)$ is the probability current,
$j_{in}$ is the incoming probability current, and $j_{r}$ is the
radial probability current. Integrating over solid angles, we find
that\begin{equation}
\sigma=\int \frac{\int j_{r}r^{2}d\Omega}{j_{in}}d\Omega\label{eq:freespxsec}\end{equation}
The optical theorem in free space relates the forward scattering amplitude
to the cross section.

Here, we modify the above free space relations, in order to define
a cross section for our confined geometry. With incident mode $\phi_{n}(x,y)$,
the scattering wavefunction is\begin{eqnarray}
\psi_{n}(x,y)-\phi_{n}(x,y) & = & -\frac{i}{\sqrt{k_{x}^{(n)}}}\sum_{m}\frac{\mathcal{R}s}{\sqrt{k_{x}^{(m)}}}\chi_{n}(y_{0})\chi_{m}(y_{0})\nonumber \\
 &  & \hspace{.5in}\times\frac{e^{ik_{x}^{(m)}|x|}}{\sqrt{k_{x}^{(m)}}}\chi_{m}(y)\label{eq:scatteringwfn}\end{eqnarray}
which yields a probability current in the $x$ direction\begin{eqnarray}
j_{x} & = & \frac{|\chi_{n}(y')|^{2}}{k_{x}^{(n)}}\sum_{m,p}|\mathcal{R}s|^{2}\frac{1}{\sqrt{k_{x}^{(n)}k_{x}^{(p)}}}\chi_{m}(y_{0})\chi_{p}(y_{0})\nonumber \\
 &  & \hspace{.5in}\times e^{i(k_{x}^{(m)}-k_{x}^{(p)})|x|}\chi_{m}(y)\chi_{p}(y).\label{eq:current}\end{eqnarray}
Integrated over $y,$ the cross terms cancel, and this current yields
a scattered flux\begin{equation}
\int_{0}^{d}j_{x}dy=\frac{\chi_{n}^{2}(y_{0})}{k_{x}^{(n)}}\sum_{m}|\mathcal{R}s|^{2}\frac{\chi_{m}^{2}(y_{0})}{k_{x}^{(m)}}\end{equation}
going in each direction. As our basis is normalized to unit flux,
in our case, the incident flux is simply 1. We define a differential
cross section for the $n^{{\rm th}}$ incoming mode, similar to the
free space differential cross section (\ref{eq:freespxsec}):\begin{eqnarray}
\sigma_{n} & = & \int dy\frac{d\sigma_{n}}{dy}\\
 & = & \int dy\frac{\int j_{x}dy}{j_{in}}\\
 & = & |\mathcal{R}s|^{2}d\frac{\chi_{n}^{2}(y_{0})}{k_{x}^{(n)}}\sum_{m}\frac{\chi_{m}^{2}(y_{0})}{k_{x}^{(m)}}\label{eq:sigmanpartial}\end{eqnarray}
where we note the exact analogy between plane waves in our waveguide
and partial waves in free space. Note also that $\sigma_{n}$ has
the correct dimensions of length.

In free space, the cross section of a spherically symmetric scatterer
is independent of the direction of the incident plane wave. In the
waveguide, however, the cross section depends on the incoming mode,
because the waveguide breaks spherical symmetry. We can define a total
cross section as the average over incoming directions,\begin{eqnarray}
\bar{\sigma} & = & \frac{1}{2N}\sum_{n=-N}^{N}\sigma_{n}\end{eqnarray}
We note that $\bar{\sigma}$ represents a sort of fraction of the
incoming wavefunctions which scatter. Flux conservation imposes the
bound\begin{equation}
0\leq\sum_{n=-N}^{N}\sigma_{n}\leq d.\end{equation}
 The maximal value of $\bar{\sigma}$ will, therefore, be $1/2N$,
and its minimal value will be 0.

We had previously mentioned that ${\rm Rank(}\mathbf{T}-\mathbf{I})={\rm Rank(\mathbf{R})=}1.$
This reduced rank indicates that, as in free space, a basis exists
in which a single wavefunction scatters, while the remaining $2N-1$
do not. The quantity $2N\bar{\sigma}$ thus tells us the fraction
of the flux of this scattering wavefunction, and should be between
0 and 1. Henceforth we will speak of $\sigma\equiv2N\bar{\sigma}/d$
as the cross section (as a fraction of the wire width). Our final
cross section is thus\begin{eqnarray}
\sigma & = & |\mathcal{R}s|^{2}\left(\sum_{n=-N}^{N}\frac{\chi_{n}^{2}(y_{0})}{k_{x}^{(n)}}\right)^{2}\label{eq:crossModes}\end{eqnarray}
and satisfies\begin{equation}
0<\sigma<1.\end{equation}
We note that for $m$ impurities in the wire, assuming there are $N>m$
modes available, ${\rm Rank(}\mathbf{T}-\mathbf{I})={\rm Rank(\mathbf{R})=}m.$
By arguments analogous to the single scatterer case, with $m$ scatterers
present, we could choose a basis where only $m$ wavefunctions would
scatter, while the remaining $N-m$ would be transmitted without any
perturbation.

\subsubsection{Modified Optical Theorem in the Waveguide\label{sub:Modified-Optical-Theorem}}

The usual free space optical theorem involves the imaginary part of
the forward scattering amplitude. One can show that the free space
optical theorem constrains $s$ via\begin{equation}
\left({\rm Im}s\right)^{2}=-\frac{1}{2}\left|s\right|^{2}.\end{equation}
A similar relation holds in the wire, and as it turns out, flux conservation
implies that not every value of $\mathcal{R}s$ is physically permissible.
Unitarity of $\textbf{S}$ implies that\begin{equation}
\sum_{m}|R_{mn}|^{2}+|T_{mn}|^{2}=1.\label{eq:unitarityCondition}\end{equation}
Using (\ref{eq:rmatrix}-\ref{eq:trmatrix}) in (\ref{eq:unitarityCondition}),
after some algebra, we find\begin{equation}
\begin{array}{cccc}
\sum_{m}|R_{mn}|^{2}+\delta_{mn}(1+2{\rm \,\, Re\,\,}R_{mn}) & = & 1, & {\rm any\,\,}n\end{array}\label{eq:rmatrixtmatrix}\end{equation}
which requires that $\mathcal{R}s$ satisfy the constraint\begin{equation}
|\mathcal{R}s|^{2}\sum_{m}\frac{\chi_{m}^{2}(y_{0})}{k_{x}^{(m)}}=-{\rm Im}(\mathcal{R}s).\label{eq:optical}\end{equation}
We can express (\ref{eq:optical}) as an optical theorem, like the
one in free space. Define the ``forward scattering amplitude''
to be the amplitude which scatters in the direction of the incoming
wave:\begin{equation}
f_{n}\equiv-\frac{i\mathcal{R}s}{k_{x}^{(n)}}\chi_{n}(y_{0}).\label{eq:fn}\end{equation}
With this definition, our ``optical theorem'' in the wire becomes\begin{equation}
{\rm Im}(e^{-i\pi/2}f_{n})=\chi_{n}(y_{0})\sigma_{n}\label{eq:opticaltheoremwirefn}\end{equation}
which is very similar to the 2-D free space optical theorem, \begin{equation}
{\rm Im}\left(f(0)e^{-i\pi/4}\right)=\sqrt{\frac{k}{8\pi}}\sigma\label{eq:opticalthmfreespace}\end{equation}
where $f(0)$ is the forward scattering amplitude. Combining (\ref{eq:crossModes})
and (\ref{eq:optical}) allows us to define a cross section,\begin{eqnarray}
\sigma(k,a) & = & -\sum_{n}\frac{\chi_{n}(y_{0})^{2}}{k_{x}^{(n)}}{\rm Im}(\mathcal{R}s)\label{eq:sigmaCalculation}\\
 & = & \frac{{\rm Im}(\mathcal{R}s)^{2}}{|\mathcal{R}s|^{2}}.\label{eq:sigmaCalculation2}\end{eqnarray}

\subsubsection{Relationship between Cross Section and Conductance\label{sub:Relationship-between-Cross}}

We note that our cross section is trivially related to the conductance
of the wire, calculated as in the Landauer formalism\cite{FisherLee}
as $G=\frac{2e^{2}}{\hbar}{\rm Tr}\;\textbf{T}^{\dagger}\textbf{T}.$
Noting that $\textbf{T}=\textbf{I-\textbf{R}},$ and letting $N$
equal the number of open channels,\begin{eqnarray}
G & \propto & {\rm Tr}\;\textbf{T}^{\dagger}\textbf{T}.\nonumber \\
 & = & N+{\rm 2Re\, Tr\,\textbf{R}}+\sum_{m}\sum_{p}|R_{mp}|^{2}\label{eq:tdt1}\\
 & = & N+2{\rm Im\,}(\mathcal{R}s)\sum_{p}\frac{\chi_{p}^{2}(y_{0})}{k_{x}^{(p)}}\label{eq:tdt2}\\
 &  & \hspace{.2in}+\sum_{p}\frac{1}{k_{x}^{(p)}}\chi_{p}^{2}(y_{0})\sum_{m}\frac{|\tilde{s}|^{2}}{k_{x}^{(p)}}\chi_{m}^{2}(y_{0})\nonumber \\
 & = & N-2\sigma-{\rm Im\,}(\mathcal{R}s)\sum_{p}\frac{\chi_{p}^{2}(y_{0})}{k_{x}^{(p)}}\label{eq:tdt3}\\
 & = & N-\sigma\label{eq:conductanceXsec}\end{eqnarray}
where we have used (\ref{eq:rmatrix}), (\ref{eq:optical}), and (\ref{eq:sigmaCalculation}).

In this section, we have laid out the formalism which we shall use
to examine scattering in the wire. In Section \ref{sec:Results-and-Discussion:},
we use our formalism to examine how the cross section and conductance
of the confined scatterer vary as the physical parameters of the wire
change.

\section{Results and Discussion: Confinement-Induced Scattering Phenomena\label{sec:Results-and-Discussion:}}

Confinement-induced effects on transport, in particular  ``conductance
dips'' as each new mode opens, have been observed experimentally
\cite{impurityExperiment1} as well as in other theoretical
investigations\cite{ImpurityImages,ImpurityTheoryBagwell,ImpurityTheory2004,ImpurityTheoryReichl}
of impurity scattering in quantum wires.  Changes in the transport
due to motion of a single defect are studied in relation to universal
conductance fluctuations.\cite{impurityExperiment2} Theoretically,
the observed phenomena are generally explained via mode-mixing effects,\cite{ImpurityTheoryBagwell}
or loosely attributed to multiple scattering from the walls.\cite{ImpurityImages}
In this section, we discuss some confinement induced phenomena as
we specifically observe them in our wire.

The first quantity we wish to examine is how the cross section of
the confined scatterer relates to the free space cross section of
the identical scatterer, as a function of energy. The cross section
of a free space scatterer is\begin{equation}
\sigma_{f}=\frac{1}{k}\left|s\right|^{2}.\label{eq:sigmaf}\end{equation}

\begin{figure}[htb]
\begin{center}\includegraphics[%
  width=.9\columnwidth,
  keepaspectratio]{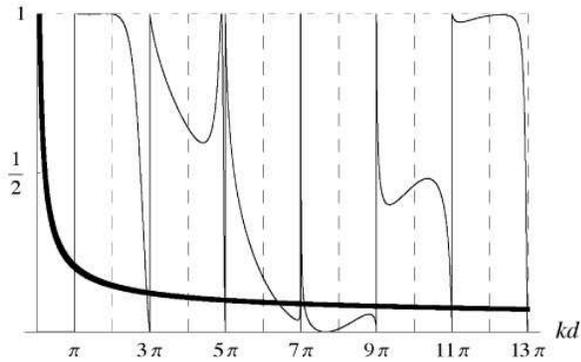}\end{center}
\caption{Free space cross section (thick line) and cross section for confined
scatterer (thin line), vs. $kd.$ The scatterer is in the center of
the wire, at $y_{0}=0.5d.$ Its scattering length is $a=0.1d.$ \label{fig:freeSpace}}
\end{figure}
Fig. \ref{fig:freeSpace} shows that one feature resulting purely
from confinement is the appearance of sharp discontinuities in the
cross section as new modes become available; the effect is present
only if the newly opened mode is nonzero at the scatterer. To the
left of each mode opening, the scatterer appears entirely transparent,
whereas to the right of the mode opening its cross section is the
full width of the wire. Another point to note is that, while the free
space cross section decreases monotonically with $k$, the cross section
of the confined impurity does not, and can be nonzero at arbitrarily
high $k$.

\subsection{Resonances}

\begin{figure*}[p]
\begin{center}\includegraphics[%
  width=1\textwidth,
  keepaspectratio]{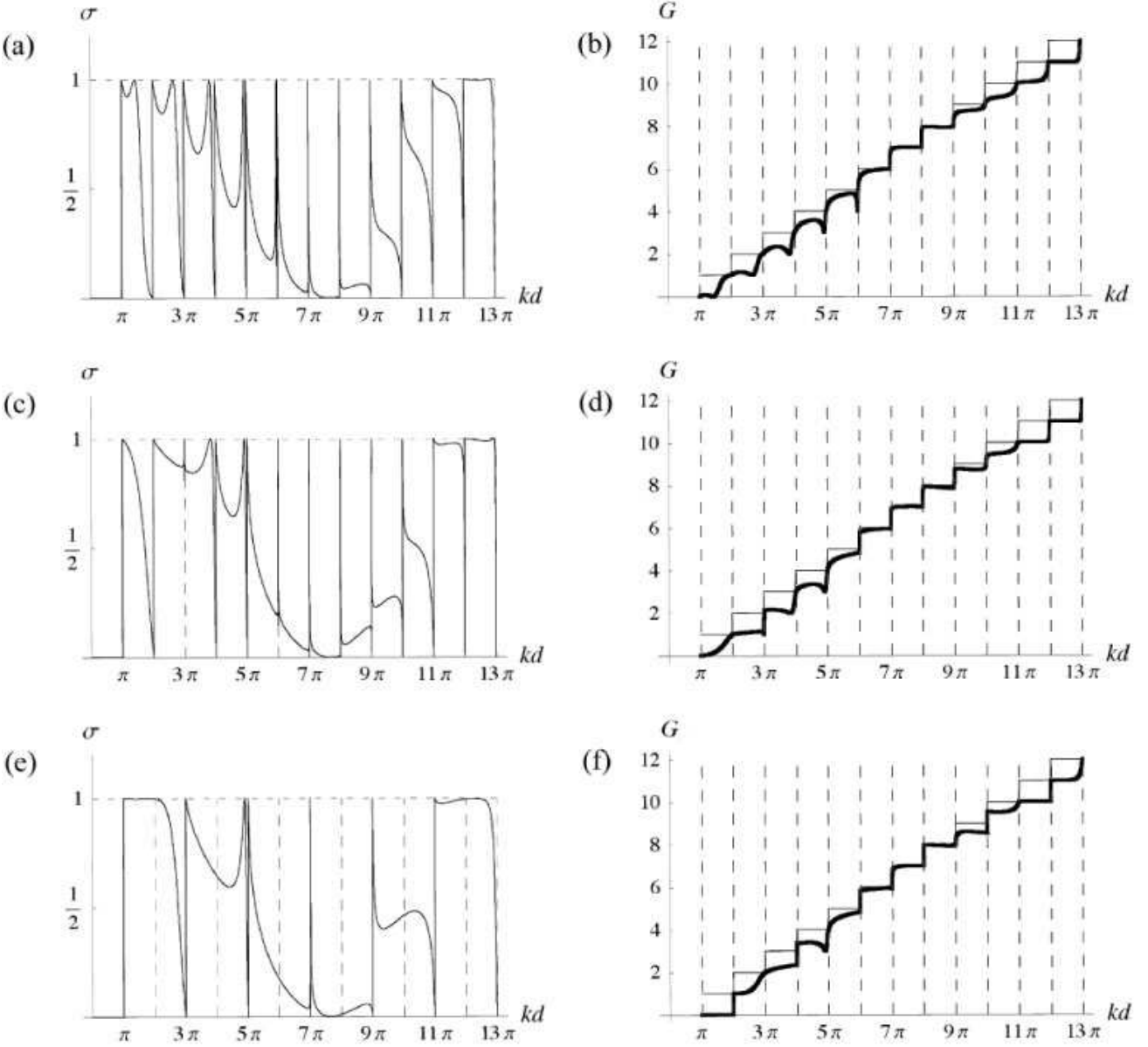}\end{center}
\begin{singlespace}
\caption{Cross sections (a,c,e)  and conductances (b,d,f) vs. the wavenumber,
$k$, for the scatterer at three different positions in the wire.
 The conductance is in units of the conductance quantum, $2e^{2}/\hbar.$
The scatterer positions are $y_{0}=0.05d$ (a, b), $y_{0}=0.32d$
(c, d), and $y_{0}=0.50d$ (e, f).    The scattering length is $a=0.1d.$
This figure illustrates resonances, and how they change with scatterer
position. Resonances in the cross section appear as new transverse
modes open $(kd=n\pi).$ In general, just below each mode opening,
$\sigma=0$.  Just above each mode opening, $\sigma$=1. By moving the scatterer around, we can tune the character of the resonances.  In (a), the scatterer is at $y_{0}=0.05$.  Since $\chi_{n}(0.05)\neq0$
for $n$ any available mode, each resonance obeys the general rules
(\ref{eq:sigmaleft},\ref{eq:sigmaright}). In (b), because $\chi_{3}(0.32d)\ll1$,
the $kd=3\pi$ resonance is very narrow, and barely there.  In (c),
with the scatterer in the precise center of the wire, $\chi_{n}(0.05d)\neq0$
where $n$ is any even available mode, and thus the resonances at
$kd=0,2\pi,4\pi,\dots$ have vanished entirely. Discontinuities in
the cross section lead to a continuously varying conductance: the
two quantities are related by (\ref{eq:conductanceXsec}). The impurity
reduces the conductance by at most one channel.  Comparing the conductance
with (heavy line) and without the impurity (thin line), we observe
that to the right of each mode opening, the conductance falls, generally
to the value of the next-lowest stairstep.  Several
references\cite{ImpurityImages,ImpurityTheoryReichl,ImpurityTheoryBagwell,ImpurityTheory2004}
obtain slightly different results for the conductance vs. wavenumber
curve, in which $\sigma=1$  just \emph{below} the
mode opening. We attribute the discrepancy to differences in how the
point scatterer is modeled. 
\label{fig:xseccond}}
\end{singlespace}
\end{figure*}

We wish to examine some of the features of the confined cross section
in Fig. \ref{fig:freeSpace} in greater detail. In particular, we
want to look at the resonances, and understand the effect of varying
the scatterer position. Fig. \ref{fig:xseccond} %
shows how the cross section and conductance vary as functions of
energy, for three different values of $y_{0}$. We observe the general
property that resonances in the cross section appear as new transverse
modes open $(kd=n\pi),$ unless the newly opened mode is zero at the
scatterer. We further observe that the resonances have a universal
structure: If a resonance exists at $kd=n\pi$, then\begin{eqnarray}
\lim_{kd\rightarrow n\pi^{-}}\sigma(k) & = & 0\label{eq:sigmaleft}\\
\lim_{kd\rightarrow n\pi^{+}}\sigma(k) & = & 1.\label{eq:sigmaright}\end{eqnarray}

As the scatterer position changes, the character, shape, and number
of resonances change also. In Fig. \ref{fig:xseccond}a), the scatterer
is located at $y_{0}=0.05d$, where each of the modes is nonzero.
Each resonance obeys (\ref{eq:sigmaleft}-\ref{eq:sigmaright}). In
Fig. \ref{fig:xseccond}c), the scatterer position is $y_{0}=0.32d.$
Because $\chi_{3}(0.32d)\ll1$, the resonance at $kd=3\pi$ is much
narrower than the others, and will in fact vanish if we move the scatterer
slightly to $y_{0}=d/3$. In Fig \ref{fig:xseccond}e), we observe
such missing resonances--with the scatterer in the precise center
of the wire, all even modes are zero there, and the resonances at
$kd=2\pi,4\pi,6\pi,\dots$ have vanished.

Figs. \ref{fig:xseccond}b), \ref{fig:xseccond}d), and \ref{fig:xseccond}e)
show the conductance for each scatterer position, which is related
to the cross section by (\ref{eq:conductanceXsec}). Discontinuities
such as (\ref{eq:sigmaleft}-\ref{eq:sigmaright}) in the cross section
lead to a continuously varying conductance. The impurity reduces the
conductance by at most one stairstep.  Comparing the conductance with
(heavy line) and without the impurity (thin line), we observe that
to the right of each mode opening, the conductance falls, generally
to the value of the next-lowest stairstep.

From Fig. \ref{fig:xseccond}, our observations are as follows: (1)
Resonances occur for values of $k$ at which a new mode opens, unless
the new mode is zero at the scatterer. (2) The general structure of
a resonance is that the scatterer becomes transparent immediately
before a mode opens, and that its cross section jumps to one immediately
after the mode has opened. (3) The shape and width of the resonances
depend on the scatterer position, $y_{0}.$ In Section \ref{sec:Physical},
we shall prove these statements, and explain them in terms of interference
effects.

Although we have examined the case of a repulsive
scatterer in Fig. \ref{fig:xseccond}, our analytic
and numerical results indicate that even for an attractive impurity
($a<0$) the conductance is reduced by a single unit immediately after
the subband opening $kd=n\pi$, rather than just below it. Similar
results about the conductance reduction near the opening of each subband
appear in other theoretical investigations of quantum wires; we cite some representative works.\cite{ImpurityImages,ImpurityTheory2004,ImpurityTheoryBagwell,ImpurityTheoryReichl}
One result\cite{ImpurityImages} is for a scatterer in a rectangle, and therefore
cannot be exactly compared to ours. However, the remaining references\cite{ImpurityTheoryReichl,ImpurityTheoryBagwell,ImpurityTheory2004}
treat the infinite wire, and show a slight difference from ours: In
Refs. 6 and 7, which examine only attractive impurities, the reduction
in conductance appears slightly \emph{below} the
subband opening. Ref. 10 examines both attractive and repulsive scatterers.
For repulsive scatterers, Ref. 10 presents results similar to our
Fig. \ref{fig:xseccond}, while for attractive scatterers
the results of Ref. 10 resemble those in Refs. 6 and 7. Because delta functions
in more than one dimension do not scatter, many different methods
exist of representing a point impurity in two dimensions. We have chosen
to use a $t$ matrix representation as in Refs. 12, 16, and 17, and the particular
$t$ matrix we have chosen in (\ref{eq:optical1}) mimics the $s$ wave scattering of a hard
disk of radius $a$.  The references whose results differ from ours
have used explicit limiting forms of delta potentials. We attribute the slight difference in results to
the different forms and representations of the point impurity.

\subsection{Between Resonances}

\begin{figure}[htb]
\begin{center}\includegraphics[%
  width=.9\columnwidth,
  keepaspectratio]{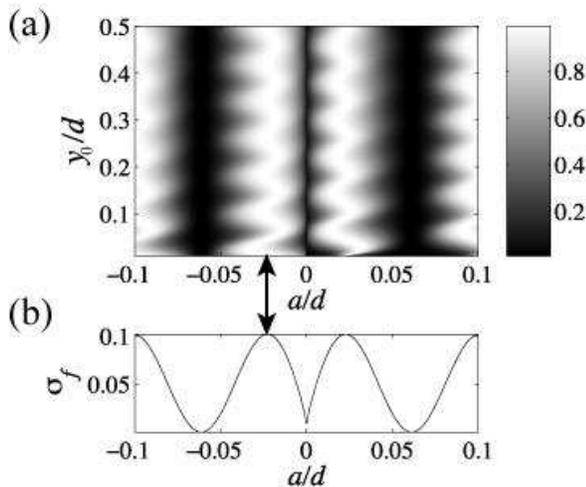}\end{center}
\caption{(a) Cross section of the confined scatterer for $kd=12.5\pi$ (halfway
between having 12 and 13 modes open). On the horizontal axis, we vary
the effective scattering radius $a$ from $-0.1d$ to $0.1d$. The
vertical axis is $y_{0},$ the vertical position of the scatterer
in the waveguide: $y_{0}$ varies from the edge of the wire ($y_{0}=0)$
to the middle of the wire ($y_{0}=0.5$). In contrast to the case
of a resonance, between resonances, the cross section is almost identical
to the free space cross section $\sigma_{f}$, shown in (b); the arrow
indicates two corresponding maxima. The only effect of varying $y_{0}$
is to introduce some small oscillations due to interference effects.
The plots continue in a similar manner to the left and right.\label{fig:crossSection}}
\end{figure}

We have observed that the scatterer position in the wire influences
not only the width and shape of the resonances, but can even cause
a resonance to disappear. Between resonances, we note the opposite
phenomenon: The scatterer position barely affects the cross section
at all. As we vary the scattering length $a$, the cross section behaves
very similarly to the free space cross section (see Fig. \ref{fig:crossSection}).

Our general observations from the numerical results are the following:
(1) Scattering resonances occur where nonzero modes open. (2) The
cross section jumps from zero to $d$ across a resonance. (3) The
width and existence of a resonance is influenced by $y_{0}.$ (4)
Away from resonances, the cross section behaves like the free space
cross section.

\section{Physical Phenomena Explained in Terms of a Single Scattering Wave
Function\label{sec:Physical}}

We have shown that items \ref{enu:newWf}-\ref{enu:renorm} in our
list of Section \ref{sub:freeSpaceMultiple} are characteristics of
free space scattering which have analogs in the wire. We have yet
to examine whether item \ref{enu:specialWf} has an analog as well.
In this section, we show that in the waveguide, as in free space,
a basis exists in which only a single wavefunction scatters. We write
this basis explicitly, and show that the scattering of this single
wavefunction allows us to explain the confinement-induced phenomena
shown in Figs. (\ref{fig:freeSpace}-\ref{fig:crossSection}) simply,
in terms of reflections from the scatterer and its images.

\subsection{Single Scattering Wavefunction and the Hall of Mirrors\label{sub:Single-Scattering-Wavefunction}}

In the basis of plane waves, the scatterer couples the channels so
that, in general, an incoming mode scatters into all the modes. However,

\begin{equation}
{\rm rank}\left[\textbf{S}-\left(\begin{array}{cc}
{\bf \textbf{0}} & \textbf{I}\\
\textbf{I} & \textbf{0}\end{array}\right)\right]=1,\label{eq:ranks}\end{equation}
which implies that a particular choice of basis exists in which only
$one$ of the incoming wavefunctions scatters at all, and is decoupled
from the other $N-1$ basis functions. This wavefunction is the analog
of the $s$ wave in free space.

Another way of seeing that only one wavefunction scatters is by the
following argument: For a given energy, $N$ channels are open. Any
incoming wavefunction is a linear combination of the $N$ basis functions\begin{equation}
\phi^{(n)}(x,y)=\frac{e^{\pm ik_{x}^{(n)}x}}{\sqrt{k_{x}^{(n)}}}\chi_{n}(y),\label{eq:incLR}\end{equation}
(we require only the left- or right-moving set). Any linear combination\begin{equation}
\sum_{n=1}^{N}c_{n}\phi^{(n)}(x,y)\label{eq:basisexpansion}\end{equation}
 of the $N$ basis functions (\ref{eq:incLR}) which is nonzero at
the scatterer will satisfy\begin{equation}
\sum_{n=1}^{N}c_{n}\phi^{(n)}(x_{0},y_{0})\neq0.\label{eq:wavefunction}\end{equation}
Equivalently, define an $1\times N$ matrix $\textbf{W}$ such that
$W_{n}=\phi^{(n)}(x_{0},y_{0})$ and a vector of coefficients $\vec{c}=[c_{1},\dots c_{N}].$
Then, for any scattering wavefunction, the $c_{n}$ will satisfy \begin{equation}
\textbf{\textbf{W}}\vec{c}\neq\vec{0}\label{eq:wceq0}\end{equation}
As $\textbf{W}$ is a rank 1 matrix, only one solution exists. That
solution is simply $\vec{c}\in{\rm Nul(\textbf{W})^{\perp}=Row(\textbf{W})=\textbf{W}^{\dagger}.}$
That is, the single scattering wavefunction is\begin{eqnarray}
\phi_{s}^{\rightleftharpoons}(x,y) & = & \sum_{n=1}^{N}\left(\phi^{(n)}(x_{0},y_{0})\right)^{*}\phi^{(n)}(x,y)\\
 & = & \sum_{n=1}^{N}\frac{1}{k_{x}^{(n)}}\chi_{n}(y_{0})\chi_{n}(y)e^{\pm ik_{x}^{(n)}(x-x_{0})}\label{eq:swfbothways}\end{eqnarray}
where the choice of sign determines whether the incoming wave is right
or left-moving. Suppose we take a linear combination of the two so
the scattering wavefunction is symmetric in $x.$ The scattering wavefunction
becomes\begin{equation}
\phi_{s}(x,y)=\sum_{n=1}^{N}\frac{1}{k_{x}^{(n)}}\chi_{n}(y)\chi_{n}(y_{0})\cos\left[k_{x}^{(n)}(x-x_{0})\right]\label{eq:swfplane}\end{equation}
(the other linear combination doesn't scatter). By comparison with
(\ref{eq:dattagreenswire}), we recognize the above form of the scattering
wavefunction as\begin{equation}
\phi_{s}(x,y)=-{\rm Im}G_{w}(x,y;x_{0},y_{0};k).\label{eq:spEqImG}\end{equation}
Note that all the definitions in this section, and in particular (\ref{eq:spEqImG}),
are independent of the specific form of $\chi_{n}(y)$, and thus apply
to arbitrary guides, not only to the hard wire. As we shall discuss
later, (\ref{eq:spEqImG}) is particularly important because it shows
that for an arbitrary guide, we can obtain the scattering wavefunction
entirely from the Green's function, which we can approximate semiclassically. 

We now return to the specific case of the hard wire, and try to understand
the scattering wavefunction physically. From (\ref{eq:wireGreens}),
an alternate form of the Green's function is\begin{eqnarray}
G_{w}(\vec{r},\vec{r}_{0},k) & = & -\frac{i}{2}\sum_{n=-\infty}^{\infty}(-1)^{n}H_{0}(k\left|\vec{r}-\vec{r}_{n}\right|)\label{eq:hankelGreens}\end{eqnarray}
where the $\vec{r}_{n}$ are the image positions. Combining (\ref{eq:spEqImG})
and (\ref{eq:hankelGreens}), we find an alternate expression for
the scattering wavefunction:\begin{equation}
\phi_{s}(x,y)=\frac{1}{2}\sum_{n=-\infty}^{\infty}(-1)^{n}J_{0}(k\left|\vec{r}-\vec{r}_{0}\right|).\label{eq:specialBessel}\end{equation}
Eq. (\ref{eq:specialBessel}) shows that the analog of the free space
$s$ wave in the hard wire is, as we might have expected, simply the
free space $s$ wave, plus an infinite series of images--a  ``hall
of mirrors $s$ wave.'' We can complete the analogy between free
space and our wire by expressing the $N-1$ unscattered wavefunctions
in terms of a {}``mirrors'' basis. We do so by including higher
order mirrored waves as in Table \ref{table:swfs}. %
\begin{table*}[htb]
\begin{center}\begin{tabular}{|c|c|>{\centering}m{3.5in}|}
\hline 
Mirrored wave&
Expression&
Wavefunction plotted for $kd=40$ (twelve modes open)\tabularnewline
\hline
$s$&
 $\frac{1}{2}\sum_{n=-\infty}^{\infty}(-1)^{n}J_{0}(k\left|\vec{r}-\vec{r}_{n}\right|)$&
\includegraphics[%
  scale=0.3]{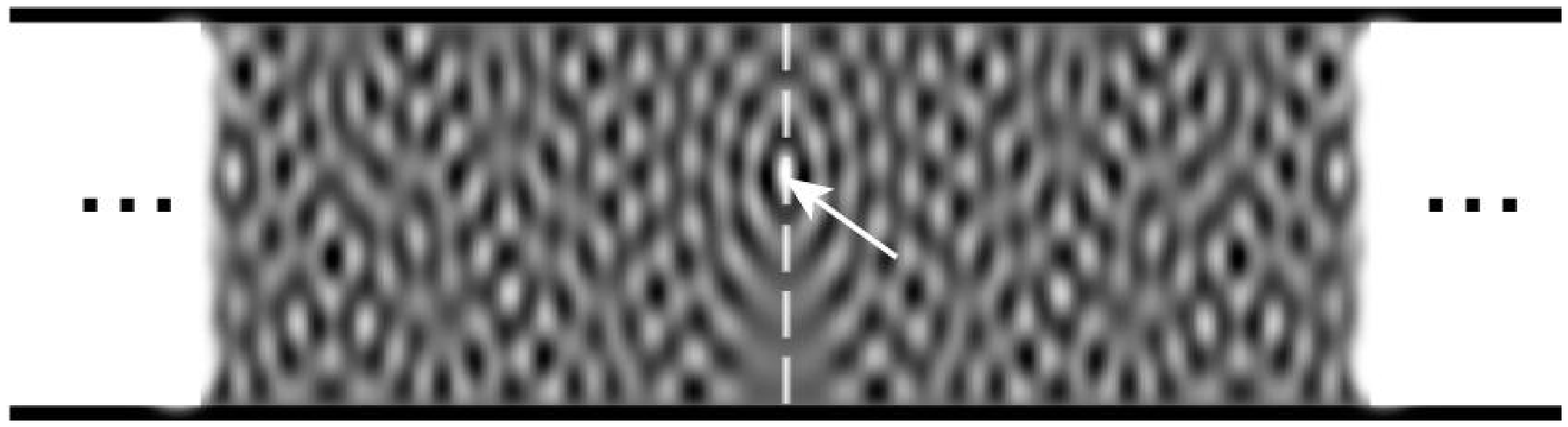}\tabularnewline
\hline
$p_{x}$&
 $\frac{1}{2}\sum_{n=-\infty}^{\infty}(-1)^{n}J_{1}(k\left|\vec{r}-\vec{r}_{n}\right|)\cos\theta$&
\includegraphics[%
  scale=0.3]{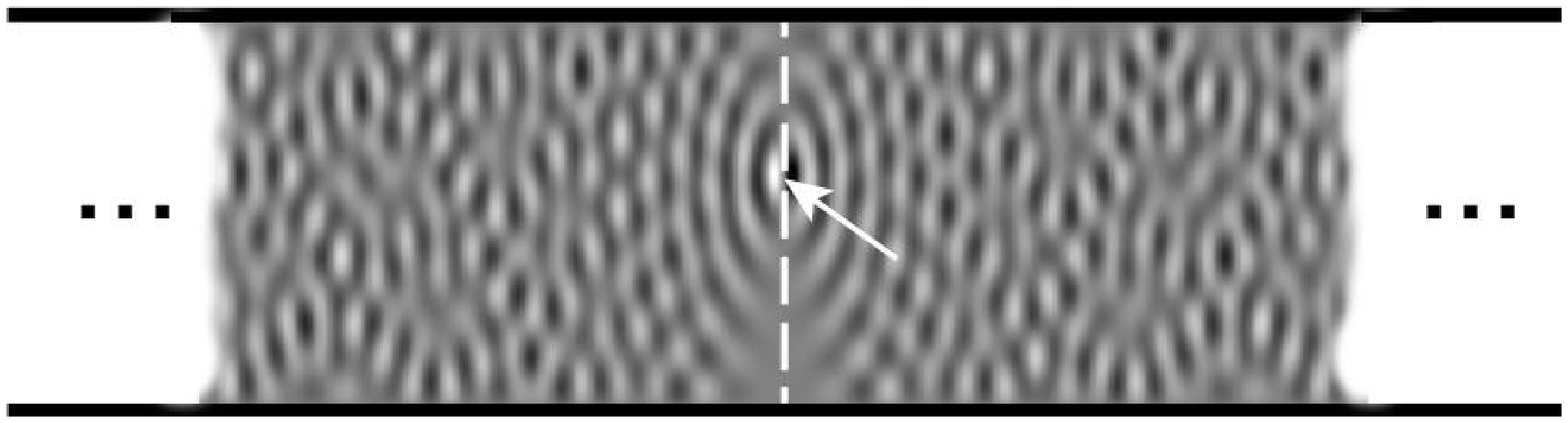}\tabularnewline
\hline
$d_{xy}$&
 $\sum_{n=-\infty}^{\infty}J_{2}(k\left|\vec{r}-\vec{r}_{n}\right|)\sin2\theta$&
\includegraphics[%
  scale=0.3]{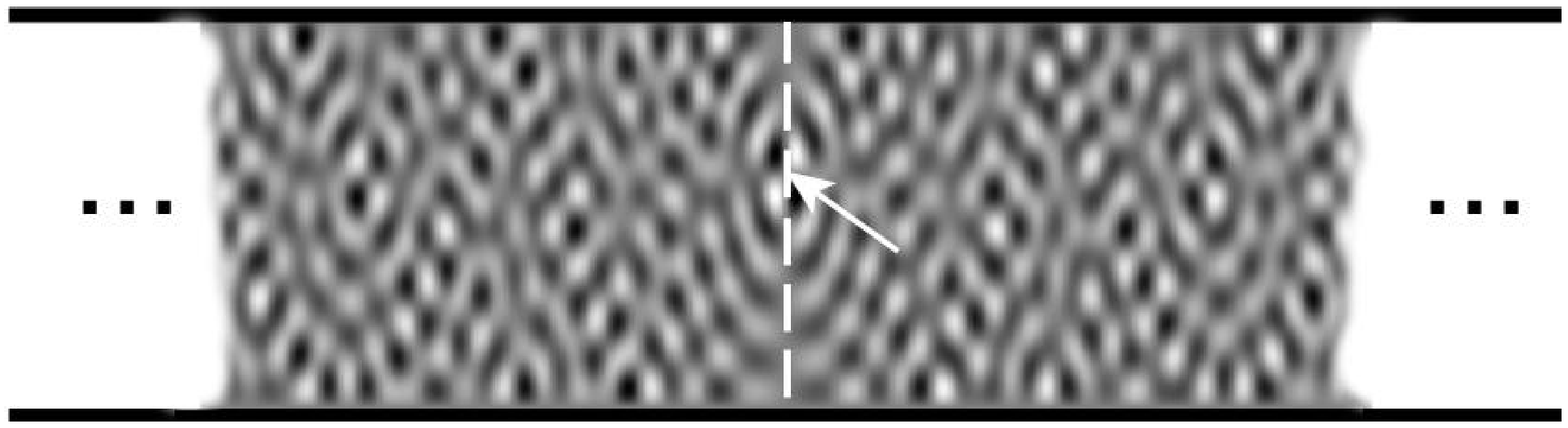}\tabularnewline
\hline
$f_{x^{3}-3xy^{2}}$&
 $\sum_{n=-\infty}^{\infty}(-1)^{n}J_{3}(k\left|\vec{r}-\vec{r}_{n}\right|)\cos3\theta$&
\includegraphics[%
  scale=0.3]{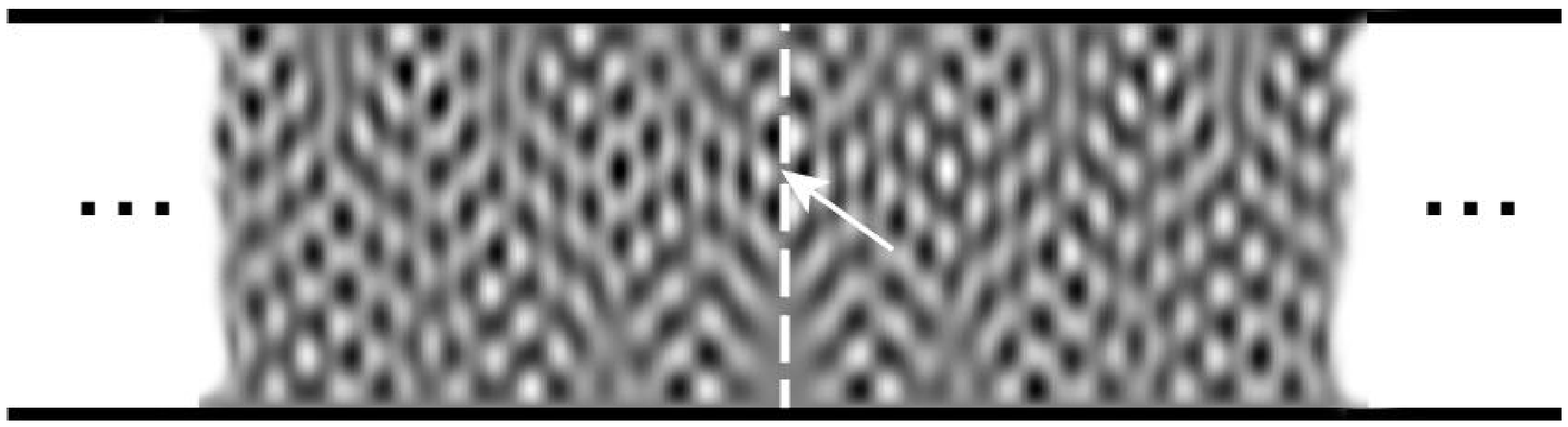}\tabularnewline
\hline
$\vdots$&
 $\vdots$&
$\vdots$\tabularnewline
\hline
\end{tabular}\end{center}

\caption{Mirrors basis. The first column indicates the symmetry, the second
column the expression, and the third column a plot, of each mirrored
wave. The arrows in the plots indicate the impurity, located at $y_{0}=0.6d$.
The mirror $s$ wave (first row) is nonzero at the impurity position
$\vec{r}_{0},$ and thus scatters. The scatterer is, however, transparent
to the higher partial waves shown in the second through fourth rows.
These partial waves, expressed as $p_{x}$,$d_{xy}$,$f_{x^{3}-3xy^{2}}$
waves plus images, do not scatter due to their nodal lines along $x=x_{0}$
(indicated by the dashed white lines in the plots). \label{table:swfs}}
\end{table*}

The mirroring, and the signs assigned to each image, are somewhat
subtle: Only those combinations of free space orbitals which satisfy
the boundary conditions, but are also zero at the scatterer, will
not scatter. We note that the basis in Table \ref{table:swfs} is
$not$ an orthogonal one; inner products of an $p$ wave centered
on one image and a $d$ wave centered on another image, for example,
do not vanish, nor do the various such cross-terms cancel each other.
However the important point is that the higher partial waves we have
defined are all orthogonal to the mirror $s$ wave: Higher mirrored
partial waves do not scatter, whereas the mirror $s$ wave does.

We note that the image sums in Table \ref{table:swfs} converge slowly
and are unsuitable for numerical purposes. However, we have the plane
wave expansion (\ref{eq:swfplane}) for the $s$ wave, as well as
the analogous expansion\begin{eqnarray}
\phi_{s}^{+}(\vec{r}) & = & \frac{1}{2}\sum_{n=-\infty}^{\infty}J_{0}(k\left|\vec{r}-\vec{r}_{n}\right|)\label{eq:besselswf}\\
 & = & \frac{2}{d}\sum_{n=1}^{N}\frac{1}{k_{x}^{(n)}}\cos\left(\frac{n\pi y}{d}\right)\cos\left(\frac{n\pi y_{0}}{d}\right)\nonumber \\
 &  & \hspace{.5in}\times\cos\left[k_{x}^{(n)}(x-x_{0})\right]\label{eq:psisplus}\end{eqnarray}
 for the $s$ wave plus $positive$ images. We can apply the raising
operator \begin{equation}
\hat{L}_{+}\equiv\frac{1}{k}\left(\partial_{x}+i\partial_{y}\right)\end{equation}
to these plane wave expansions to obtain plane wave expansions for
the higher partial waves. For example, we can obtain plane wave expansions
of the waves in Table \ref{table:swfs} as\[
\begin{array}{cclcl}
\phi_{p_{x}}(\vec{r}) & = & {\rm Re}\left[\hat{L}_{+}\phi_{s}(\vec{r})\right] & = & \frac{1}{k}\partial_{x}\phi_{s}(\vec{r})\\
\phi_{d_{xy}}(\vec{r}) & = & {\rm Im}\left[\hat{L}_{+}^{2}\phi_{s}^{+}(\vec{r})\right] & = & \frac{2}{k^{2}}\partial_{xy}\phi_{s}^{+}(\vec{r})\\
\phi_{f_{x^{3}-3xy^{2}}}(\vec{r}) & = & {\rm Re}\left[\hat{L}_{+}^{3}\phi_{s}(\vec{r})\right] & = & \frac{1}{k^{3}}\left(\partial_{x}^{3}-3\partial_{x}\partial_{y}^{2}\right)\phi_{s}(\vec{r}),\end{array}\]
and we could proceed similarly to obtain higher partial waves.

\subsection{Interference Effects\label{sub:Physical-phenomena-explained}}

We might expect that the cross section is related to the fraction
of the single scattering wavefunction which actually scatters, and
we might thus expect the cross section to be proportional to the amplitude
of the mirror wavefunction at the scatterer. However, some numerics
reveal that the mirror wavefunction alone is insufficient to fully
describe the scattering. From (\ref{eq:crossModes}) and (\ref{eq:swfplane}),
we find\begin{eqnarray}
\sigma & = & \left|\mathcal{R}s\right|^{2}\phi_{s}(\vec{r}_{0})^{2}\\
 & = & \left|s\tilde{\phi}_{s}(\vec{r}_{0})\right|^{2}\label{eq:sigmaSwf}\end{eqnarray}
meaning that the relevant quantity is the $renormalized$ mirror wavefunction,
and that its value at the scatterer alone explains all the scattering
phenomena. Eq. (\ref{eq:sigmaSwf}) is again independent of the form
of the waveguide. 

We have already discussed how renormalization of the scattering wavefunction
arises from interference between all possible bounces off images.
The implication of (\ref{eq:sigmaSwf}) is that the value of the resulting
interference pattern at the location of the scatterer determines the
cross section: At configurations (wire widths, incident wavenumbers,
etc.) for which this interference is destructive at the scatterer,
the scatterer is transparent. If the system is configured so that
the bounces interfere constructively at the scatterer, a resonance
results.

\subsubsection*{Semiclassical Approximations for General Wires}

We wish to highlight another important point, which is that combining
(\ref{eq:sigmaSwf}) with (\ref{eq:gr}) and (\ref{eq:spEqImG}),
we find that\begin{equation}
\sigma=\lim_{\vec{r}\rightarrow\vec{r}_{0}}\left\{ \left|\frac{s{\rm Im}G_{w}(\vec{r},\vec{r}_{0})}{1-s\left[G_{w}(\vec{r},\vec{r}_{0})-G_{0}(\vec{r},\vec{r}_{0})\right]}\right|^{2}\right\} .\label{eq:semiclassical}\end{equation}
The significance of (\ref{eq:semiclassical}), which applies to an
arbitrary wire, is that the cross section is fully determined by the
Green's function of the empty guide. This relation is important because
the Green's function is a quantity which we can approximate semiclassically.
Substituting the semiclassical Green's function into (\ref{eq:semiclassical})
will give us a semiclassical approximation to the cross section. Semiclassical
methods may be of use in determining the transport properties of guiding
potentials which are more complicated than the hard wire, and thus
not amenable to exact quantum treatment.

\subsection{Scattering Resonances and Conductance Reduction\label{sub:Universal-Conductance-Reduction}}

In Figs. \ref{fig:freeSpace}-\ref{fig:xseccond}, we observed a reduction
in the conductance, which are similar (although not identical) to
the conductance dips observed in other theoretical investigations.\cite{ImpurityImages,ImpurityTheoryBagwell,ImpurityTheory2004,ImpurityTheoryReichl}
Using our formalism, we can understand the resonances and conductance
dips of Figs. \ref{fig:freeSpace}-\ref{fig:xseccond} simply, in
terms of interference between different paths. Again the results of
this section do not depend on the specific form of the transverse
modes $\chi_{n}(y)$, and consequently apply to a general waveguide.

As we showed in (\ref{eq:bouncesFeynman}), the renormalized mirror
wavefunction contains contributions that have propagated from each
of the image scatterers. We are interested in the behavior of a cross
section when a mode is about to open. The mirror wavefunction involves
only open channels, and is thus discontinuous across a mode opening.
Using (\ref{eq:swfplane}), we can examine the behavior of the mirror
wavefunction near $kd=N\pi$, where the $N^{{\rm th}}$ mode opens.
Suppose that the newly opened mode is nonzero at the scatterer, $\chi_{N}(y_{0})\neq0.$
On the left of the mode opening,\begin{eqnarray}
\lim_{\epsilon\rightarrow0}\phi_{s}\left(\vec{r}_{0};k=\frac{N\pi-\epsilon}{d}\right) & = & \frac{d}{\pi}\sum_{n=1}^{N-1}\frac{\chi_{n}^{2}(y_{0})}{\sqrt{N^{2}-n^{2}}}\label{eq:mirrorleft}\end{eqnarray}
is finite. Immediately after the mode opening, \begin{eqnarray}
\lim_{\epsilon\rightarrow0}\phi_{s}\left(\vec{r}_{0};k=\frac{N\pi+\epsilon}{d}\right) & = & \frac{d}{\pi}\frac{\chi_{N}^{2}(y_{0})}{\sqrt{2N\epsilon}}\label{eq:mirrorDiverge}\end{eqnarray}
diverges as $\epsilon^{-1/2}.$ That is, the mirror wavefunction is
always finite immediately before a nonzero mode opens, and diverges
afterwards. The value of $\chi_{N}^{2}(y_{0})$ determines the width
of the resonance (and the lifetime of the corresponding quasibound
state). 

However, while the limiting behavior of the mirror wavefunction influences
the shape of the resonances, it does not fully describe their shape.
As we discussed, the general structure of a resonance is that $\sigma$
drops to zero just before the mode opens; see e.g. Fig. \ref{fig:xseccond}.
Clearly the incoming wavefunction, which includes only modes which
are already open, cannot explain this transparency. The renormalization
factor due to the wire, however, includes $all$ modes, both evanescent
and propagating--and unlike the incoming mirror wavefunction, varies
continuously across the mode opening. Using the expression (\ref{eq:kummerform})
for $G_{r}$, we find that\begin{eqnarray}
\lim_{\epsilon\rightarrow0}G_{r}\left(k=\frac{N\pi-\epsilon}{d}\right) & = & -\frac{d}{\pi}\frac{\chi_{N}^{2}(y_{0})}{\sqrt{2N\epsilon}}\label{eq:grleft}\\
\lim_{\epsilon\rightarrow0}G_{r}\left(k=\frac{N\pi+\epsilon}{d}\right) & = & -\frac{id}{\pi}\frac{\chi_{N}^{2}(y_{0})}{\sqrt{2N\epsilon}}\label{eq:grright}\end{eqnarray}
 generally. Combining (\ref{eq:renormalizedSwfGr}) with the limits
(\ref{eq:mirrorleft}-\ref{eq:grright}), and using (\ref{eq:sigmaSwf}),
we can show that\begin{eqnarray}
\lim_{\epsilon\rightarrow0}\sigma\left(k=\frac{N\pi}{d}-\epsilon\right) & =\nonumber \\
 &  & \hspace{-1in}2N\epsilon\left|\sum_{n=1}^{N-1}\frac{\left[\chi_{n}(y_{0})/\chi_{N}(y_{0})\right]^{2}}{\sqrt{N^{2}-n^{2}}}\right|^{2}.\label{eq:limitsigmaepsilon}\end{eqnarray}
 The value of $\chi_{N}(y_{0})$ controls the width and structure
of the resonance. Examining the right side of the resonance,\begin{eqnarray}
\lim_{\epsilon\rightarrow0}\sigma\left(k=\frac{N\pi}{d}+\epsilon\right) & = & \lim_{\epsilon\rightarrow0}\left|\frac{s}{1-sG_{r}}\right|^{2}\phi_{s}^{2}(\vec{r}_{0})\label{eq:sigmaright1}\\
 & \approx & \frac{\phi_{s}^{2}(\vec{r}_{0})}{\left|G_{r}\right|^{2}}\label{eq:sigmaright2}\\
 & = & 1\nonumber \end{eqnarray}
under the assumption that $\left|sG_{r}\right|\gg1.$ An important
point to note is that when $\left|G_{r}\right|\rightarrow\infty,$
as it does at a resonance, the free space properties of the scatterer
have little effect on the cross section (see for example (\ref{eq:sigmaright1})),
and so the structure of resonances is universal.

\subsubsection*{Phase Shift and Ramsauer-Townsend Effect}

We would like to note, in passing, that the transparency of the scatterer
at certain energies is reminiscent of the Ramsauer-Townsend effect
in free space. In free space, we define the $s$ wave phase shift
$\delta_{0}$ of an $s$ wave scatterer by\begin{equation}
\psi(\vec{r})=\phi(\vec{r}_{0})+\frac{e^{2i\delta_{0}}-1}{2}H_{0}^{(1)}(k\left|\vec{r}-\vec{r}_{0}\right|),\label{eq:phaseshifts}\end{equation}
Let us define it in the confined scatterer as the phase shift of $each$
$s$ wave generated from the source and the images, when the incident
wavefunction is our mirror $s$ wave:\begin{eqnarray}
\psi(\vec{r}) & = & \phi_{s}(\vec{r})+\sum_{n=-\infty}^{\infty}\frac{e^{2i\delta_{0}}-1}{2}\label{eq:morephaseshifts}\\
 &  & \hspace{.25in}\times(-1)^{n}H_{0}^{(1)}(k\left|\vec{r}-\vec{r}_{n}\right|)\nonumber \end{eqnarray}
Comparing to the form (\ref{eq:msWf}) of the Lippmann-Schwinger equation,
we find that\begin{equation}
e^{2i\delta_{0}}=1-is\tilde{\phi}_{s}(\vec{r}_{0}).\label{eq:yetmorephaseshifts}\end{equation}
Applying (\ref{eq:sigmaSwf}), we find that\begin{equation}
\sigma=\frac{1}{4}\left|1-e^{2i\delta_{0}}\right|^{2}\label{eq:rteffect}\end{equation}
so that $\sigma=0$ when $\delta_{0}=0,\pi$ and $\sigma=1$ when
$\delta_{0}=\frac{\pi}{2},\frac{3\pi}{2}.$ As in the free space Ramsauer
Townsend effect, the cross section vanishes when our analog of the
$s$ wave phase shift does.

In conclusion, we can make the following general statements about
the cross section:

\begin{enumerate}
\item If $\chi_{N}(y_{0})\neq0,$ the cross section will be discontinuous
at $kd=N\pi.$ The limit on the left hand side will be finite, and
depend the value of $\chi_{N}(y_{0}).$ The limit on the right hand
side will be unity.
\item The width and shape of the resonances in item (2) will depend on $\chi_{N}(y_{0})$.
Physically this means that one can tune the resonances by sliding
the scatterer up and down in the wire, or by changing the nature of
the confining potential.
\end{enumerate}

\subsection{Semiclassical Interpretation of Resonances}

Having explored the structure of the resonances, we wish to understand
them semiclassically, from an interference point of view. We again
turn to the hard wire for insight into the scattering processes. The
scattering cross section depends only on a single quantity, the renormalized
mirror wavefunction. We have previously expressed this renormalized
mirror wavefunction as a sum over the different paths ending on the
scatterer (\ref{eq:bouncesFeynman}). When the interference between
these paths is maximally constructive, the cross section is maximal.
When it is maximally destructive, the cross section vanishes.

In order to make this interference clearer, we shall make a simple
approximation. Consider the most straightforward case, with the scatterer
in the center of the wire (the case we examined in Fig. \ref{fig:freeSpace}).
The image positions are $\vec{r}_{n}=(0,nd).$ In the Green's function
(\ref{eq:wireGreens}), we replace each Hankel function by its asymptotic
form, which is equivalent to making a semiclassical approximation.
This yields an approximate Green's function\begin{eqnarray}
G_{w}(\vec{r},\vec{r}_{0}) & = & \frac{1}{\sqrt{2\pi}}e^{5\pi i/4}\sum_{n=-\infty}^{\infty}\frac{1}{\sqrt{k\left|\vec{r}-\vec{r}_{n}\right|}}\label{eq:greensapprox}\\
 &  & \hspace{.25in}\times e^{i(k\left|\vec{r}-\vec{r}_{n}\right|-n\pi)}.\nonumber \end{eqnarray}
The contribution from each image carries a phase related to the path
length from the image to the observation point, as well as a Maslov
index $e^{in\pi}$ describing the sign change after $n$ reflections
from the wire walls. In the limit that $\vec{r}\rightarrow\vec{r}_{0}$,
we see that the contribution propagating to the scatterer from the
$n^{{\rm th}}$ image has relative phase $e^{i(kd-\pi)n}$. For values
of $k$ such that\begin{equation}
kd=(2p-1)\pi,\,\, p\,\,{\rm integer},\label{eq:kvalues}\end{equation}
all the scattered wavelets thus interfere constructively. Comparing
with Fig. \ref{fig:freeSpace}, we see that this constructive interference
coincides precisely with the resonances which occur as new modes open.

To fully explain the structure of the resonances, and understand the
scatterer's transparency just before modes open, we would have to
consider the effects of renormalization in this approximation. We
shall not pursue the semiclassical limit further here, as we have
already examined the exact case in great detail in Section \ref{sub:Universal-Conductance-Reduction}.
However, purely from this simple semiclassical argument, we can see
that the resonances induced by the wire are indeed related to interference
effects between different reflections.

\section{Conclusions and Future Directions}

Combining the method of images with the idea of a single scattering
wavefunction, we have developed a formalism with which to treat scattering
from a single impurity confined in a quantum guide. We find many similarities
between scattering in the confined geometry and scattering in free
space: We have defined meaningful analogs of the free space cross
section, optical theorem, partial waves, and Ramsauer-Townsend effect.
Although the hard guide is particularly useful for insight, many of
our results apply to other types of waveguides also.

We have examined the transport properties of a confined scatterer,
making several general observations about the existence, locations,
and and characters of the resonances. Additionally, we have outlined
a method for approximating the cross section and conductance semiclassically,
for arbitrary guides. We have derived an effective optical theorem
for general quantum wire potentials. Using the fact that a confined
target is rank one, we have described the transport properties of
the wire entirely in terms of a single scattering wavefunction, renormalized
by reflections from the confining potential. Our central result is
that the cross section of a confined impurity is\begin{equation}
\sigma=\left|s\tilde{\phi}_{s}\left(\vec{r}_{0}\right)\right|^{2},\label{eq:conclusion}\end{equation}
where $\tilde{\phi}_{s}\left(\vec{r}\right)$ is the renormalized
single scattering wavefunction, which includes interference effects
due to reflections from the waveguide walls.

Despite the common use of the hard-walled guide as
a theoretical model for 2DEG quantum wires,\cite{ImpurityTheory1,ImpurityTheory2004,ImpurityTheoryAJP,ImpurityTheoryBagwell,ImpurityTheoryReichl,ImpurityTheoryReichl2}
actual confining potentials for both atom waveguides as well as 2DEG
quantum wires tend to be soft or even parabolic--whereas in nanotube
quantum wires, the relevant boundary conditions are periodic (the
periodic boundary case is also exactly solvable via the method of
images, although we have not presented the derivation here). While
the hard wall itself is not the best model for 2DEGs, it is the
ideal system in which to examine how effects like conductance reduction,
which are independent of the confining potential and occur in more
realistic geometries, arise from simple reflection and interference
phenomena. The effects of multiple scattering from the impurity are
in fact very similar for general confining potentials (see e.g. the
renormalized t matrix formalism\cite{AdamThesis}).

A second limitation of this work includes the inadequacy
of the $s$ wave scatterer model at high energies, and in particular
at energies which are sometimes relevant when imaging electron flow.
In a future publication,\cite{UsForthcoming} we plan to extend our
formalism to treat higher partial waves. Another possible extension
of this work could be to examine the transport properties of many
scatterers in the wire; for example, explicitly explaining the transition
from ballistic to diffusive transport in terms of interference between
multiple mirror wavefunctions.

\begin{acknowledgments}
JYV was funded during the earliest stages of this work by a fellowship
from the National Defense Science and Engineering Graduate (NDSEG)
fellowship program. Subsequent research was funded by a grant from
the National Science Foundation, NSF CHE-0073544, and by the Nanoscale
Science and Engineering Center (NSEC) under NSF grant PHY-0117795.
JYV is grateful to M. Stopa for a helpful critical reading of the
manuscript.
\end{acknowledgments}
\bibliographystyle{apsrev}

\appendix

\section{Wire as a Diffraction Grating\label{sec:diffraction}}

In Sec. \ref{sec:greensFunction}, we used the mathematical equivalence
of the wire to a periodic array of image scatterers. Further extending
the analogy, we now reformulate the problem of the confined impurity
as one of diffraction from an infinite, periodic grating, and the
scattered wavefunction as a diffracted beam. We wish to note the detail
that because of our Dirichlet boundary conditions, our incident beam
is a sum of two plane waves rather than a single plane wave:\begin{eqnarray}
\phi_{m}(x,y) & = & \frac{1}{\sqrt{k_{x}^{(m)}}}e^{ik_{x}^{(m)}\left|x-x_{0}\right|}\chi_{m}(y)\label{eq:a1}\\
 & = & \frac{1}{id}\frac{1}{\sqrt{k_{x}^{(m)}}}e^{ik_{x}^{(m)}\left|x-x_{0}\right|}\label{eq:a2}\\
 &  & \times\left(e^{ik_{y}^{(m)}y}-e^{-ik_{y}^{(m)}y}\right).\nonumber \end{eqnarray}
Due to the linearity of Schrodinger's equation, we can consider each
of the constituent plane waves as scattering independently. The relevant
physics thus reduces to diffraction of a single plane wave incident
on a periodic array of scatterers.

\begin{figure}[htb]
\begin{center}\includegraphics[%
  width=.9\columnwidth,
  keepaspectratio]{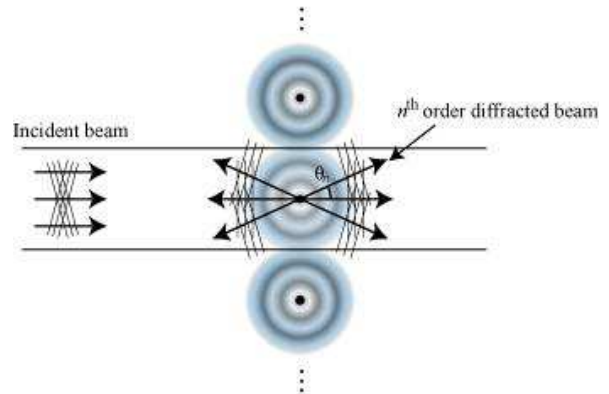}\end{center}

\caption{A mode incident on an impurity in the hard wire is, effectively,
a superposition of two plane waves incident on an infinite diffraction
grating. At large distances from the scatterer, the scattered wave
must therefore be a sum of plane waves at the Bragg angles. Closer
to the scatterers, evanescent modes appear also. \label{fig:diffractedBeam}}
\end{figure}

When our incident wavefunction strikes the diffraction grating, we
expect that in the Fraunhofer limit, far from the array, the diffracted
beam will be a sum of plane waves at the Bragg angles (see Fig. \ref{fig:diffractedBeam}).

 Each order of the diffracted beam will have a different weighting;
our object is to cast our Green's function (\ref{eq:wireGreens})
in a manner that makes these weightings explicit. This formulation
is of interest because it is equivalent to determining the transmission
and reflection coefficients of our system. Namely, if we send in a
mode, we would like to know what modes come out and with what intensities,
which is precisely what we shall find by casting our scattering wavefunction
as a diffracted beam.

To begin, we note that Bessel functions are regular, and consequently
have plane wave expansions. In fact, the well-known integral form
of the Bessel function\begin{eqnarray}
J_{0}(kr) & = & \frac{1}{2\pi}\int_{0}^{2\pi}e^{i\vec{k}(\theta)\cdot\vec{r}}d\theta\label{eq:intbessel}\end{eqnarray}
expresses the Bessel function as an isotropic sum of plane waves emanating
in all directions from the origin. We can rewrite (\ref{eq:intbessel})
in terms of a line integral over $k_{y}$as\begin{eqnarray}
J_{0}(kr) & = & \frac{1}{2\pi}\int_{C}\frac{d(k\sin\theta)}{k\cos\theta}e^{i(k_{x},k_{y})\cdot(x,y)}\\
 & = & \frac{1}{2\pi}\int_{C}\frac{dk_{y}}{k_{x}}e^{i(k_{x},k_{y})\cdot(x,y)}\label{eq:intbessel2}\end{eqnarray}
where $C$ lies along the real axis, going back and forth within the
interval $(-k,k)$.

We seek an analogous expression for the Hankel function. Since $|k_{y}|$
is always less than $k$ in (\ref{eq:intbessel2}), $k_{x}$ is always
real, and the plane waves in (\ref{eq:intbessel2}) do not include
evanescent waves. Suppose we modify the right-hand side of (\ref{eq:intbessel2}).
By permitting $k_{y}$ to range from $-\infty$ to $\infty$, we may
include contributions from all possible evanescent waves, yielding\begin{equation}
\frac{1}{2\pi}\int_{-\infty}^{\infty}\frac{dk_{y}}{k_{x}}e^{i(k_{x},k_{y})\cdot(x,y)}.\label{eq:evanescent_Sum}\end{equation}
 For $|k_{y}|>k,$ the corresponding value of $k_{x}=\pm i\kappa_{x}$,
where $\kappa_{x}\equiv\sqrt{k_{y}^{2}-k^{2}},$ is imaginary. Choosing
the positive value makes the evanescent waves die off to the right
$(x>0)$ and diverge to the left $(x<0).$ Therefore, with this choice
of sign, we may expect (\ref{eq:evanescent_Sum}) to converge only
in the right half-plane.

By adding on an evanescent contribution to the Bessel function's plane
wave expansion, we have in fact arrived (up to a constant) at an integral
form of the Hankel function, valid in the right half-plane only,

\begin{eqnarray}
H_{0}^{(1)}(kr) & = & \frac{1}{\pi}\int_{-\infty}^{\infty}\frac{dk_{y}}{k_{x}}e^{i(k_{x},k_{y})\cdot(x,y)}\label{eq:inthankel}\end{eqnarray}
where\begin{equation}
k_{x}=\left\{ \begin{array}{ccc}
\sqrt{k^{2}-k_{y}^{2}} &  & \left|k_{y}\right|\leq k\\
i\sqrt{k_{y}^{2}-k^{2}} &  & \left|k_{y}\right|>k.\end{array}\right.\end{equation}
Eq. (\ref{eq:inthankel}) is a well-known expansion,\cite{MorseFeshbach}
readily obtained by Fourier transforming Green's equation.

Consider now the sum\begin{equation}
G_{p}(\vec{r};d)=-\frac{i}{2}\sum_{n=-\infty}^{\infty}H_{0}(k\left|\vec{r}-(x_{0},nd)\right|)\label{eq:althanksum}\end{equation}
which is the Green's function for a fully periodic array of scatterers
at $(x_{0},nd)$ (or equivalently, a periodic wire with walls at $y=0,y=2d$
and a scatterer halfway between). Based on the physical picture of
diffraction (Fig. \ref{fig:diffractedBeam}), the sum in (\ref{eq:althanksum})
must be a superposition of plane and evanescent waves at real and
complex Bragg angles respectively. We wish to find the weighting on
each wave (analogous to the structure factor in X-ray diffraction).

To find the weights, we calculate the expansion explicitly. Using
(\ref{eq:inthankel}) in (\ref{eq:althanksum}), we find

\begin{eqnarray}
G_{p}(\vec{r},\vec{0}) & = & -\frac{i}{2}\sum_{n=-\infty}^{\infty}H_{0}(k|\vec{r}-\vec{r}_{n}|)\label{eq:gp1}\\
 & = & -\frac{i}{2\pi}\sum_{n=-\infty}^{\infty}\int_{-\infty}^{\infty}\frac{dk_{y}}{k_{x}}\label{eq:gp2}\\
 &  & \hspace{.5in}\times e^{i(k_{x},k_{y})\cdot(x-x_{0},y-nd)}\nonumber \\
 & = & -\frac{i}{2\pi}\int_{-\infty}^{\infty}\frac{dk_{y}}{k_{x}}e^{i(k_{x},k_{y})\cdot(x-x_{0},y)}\label{eq:gp3}\\
 &  & \hspace{.5in}\times\sum_{n=-\infty}^{\infty}\left(e^{-i(k_{y}d)}\right)^{n}\nonumber \\
 & = & -\frac{i}{d}\int_{-\infty}^{\infty}\frac{dk_{y}}{k_{x}}e^{i(k_{x},k_{y})\cdot(x-x_{0},y)}\label{eq:poisson}\\
 &  & \hspace{.5in}\times\sum_{m=-\infty}^{\infty}\delta(k_{y}-\frac{2m\pi}{d})\nonumber \end{eqnarray}
where we have used\begin{equation}
\sum_{n=-\infty}^{\infty}\left[e^{ik_{y}d}\right]^{n}=2\pi\sum_{n=-\infty}^{\infty}\delta(k_{y}d-2\pi n),\end{equation}
(similar to stationary phase) to reach the expression (\ref{eq:poisson}).
One can, more rigorously, reach (\ref{eq:poisson}) via the Poisson
sum formula.\cite{AbramowitzStegun,HankelSum}

Performing the integral in (\ref{eq:poisson}) yields the final expression\begin{eqnarray}
G_{p}(\vec{r},\vec{r}_{m};k,d) & = & -\frac{i}{d}\sum_{n=-\infty}^{\infty}\frac{1}{k_{x}^{(n)}}e^{ik_{x}^{(n)}|x-x_{0}|}\label{eq:gpfirst}\\
 &  & \hspace{.5in}\times\cos(k_{y}^{(n)}y)\nonumber \end{eqnarray}
where we use the absolute value signs to extend the Green's function
to converge on $x<x_{0},$ and we define\begin{eqnarray}
k_{x}^{(n)} & = & k\cos\theta_{n}\\
k_{y}^{(n)} & = & k\sin\theta_{n}\end{eqnarray}
 where the $\theta_{n}$ are the Bragg angles\begin{equation}
\theta_{n}=\arcsin\frac{2n\pi}{kd}\qquad n\mathrm{\: integer}\end{equation}
and the complex-valued $\theta_{n}$ are the result of the logarithmic
singularity. 

\begin{figure}[htb]
\begin{center}\includegraphics[%
  width=1\columnwidth,
  keepaspectratio]{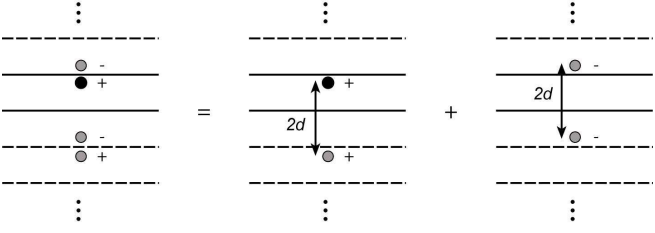}\end{center}

\caption{In the case of the off-center scatterer, the relevant image array
is the sum of two periodic arrays of opposite signs. If the source
is at $(x_{0},y_{0}),$ the image locations are $(x_{0},y_{0}+2nd$),
$(x_{0},-y_{0}+2nd$) for the positive and negative arrays respectively,
where $n=\pm1,\pm2,\dots$.\label{fig:twoArrays}}
\end{figure}

The image representation of our wire is not exactly the periodic array
of Fig. \ref{fig:diffractedBeam}. The images alternate in sign, and
in general the scatterer is off-center. We can, however, represent
our image array as a sum of two arrays, as in Fig. \ref{fig:twoArrays}.
 One can easily verify that the Green's function for the wire with
point source at $\vec{r}_{0}=(x_{0},y_{0})$ becomes the sum\begin{eqnarray}
G_{w}(\vec{r},\vec{r}_{0}) & = & \frac{1}{2}[G_{p}(\vec{r}-y_{0}\hat{y};2d)\label{eq:gwsum}\\
 &  & \hspace{.5in}-G_{p}(\vec{r}+y_{0}\hat{y};2d)]\nonumber \\
 & = & -i\sum_{n=1}^{\infty}\frac{1}{k_{x}^{(n)}}e^{ik_{x}^{(n)}|x-x_{0}|}\label{eq:imagegreensequal}\\
 &  & \hspace{.5in}\times\chi_{n}(y)\chi_{n}(y_{0})\nonumber \end{eqnarray}
where the Bragg angles are modified slightly from the case of the
periodic wire, and defined by\begin{equation}
\sin\theta_{n}=\frac{n\pi}{kd}.\end{equation}
The form (\ref{eq:imagegreensequal}) is identical to (\ref{eq:dattagreenswire}).
We have shown that diffraction translates the method of images into
the usual eigenfunction expansion of the Green's function.

\section{Improving convergence of the green's function\label{sec:kummer}}

Form (\ref{eq:dattagreenswire}) of the Green's function converges
slowly also. The reason for the slow convergence is the singularity
at the source, which is present but disguised in the evanescent modes.
We use the Kummer method for convergence acceleration,\cite{AbramowitzStegun}
which also makes the singularity explicit. Kummer's method involves
adding and subtracting a multiple of the static ($k=0$) Green's function.
We note that [12] obtains similar results, via a slightly different
procedure.

The wire Green's function is

\begin{eqnarray}
G_{w}(\vec{r},\vec{r};k) & = & -i\sum_{m=1}^{\infty}\frac{1}{k_{x}^{(m)}}\chi_{m}(y)\chi_{m}(y_{0})\label{eq:b1}\\
 &  & \hspace{.5in}\times e^{ik_{x}^{(m)}\left|x-x_{0}\right|}\nonumber \end{eqnarray}
 We wish to consider a more rapidly converging form of the Green's
function:\begin{eqnarray}
G_{w}(\vec{r},\vec{r}_{0};k) & = & \left[G_{w}(\vec{r},\vec{r}_{0};k)-\alpha G_{w}(\vec{r},\vec{r}_{0};0)\right]\label{eq:quickconverge}\\
 &  & \hspace{.5in}+\alpha G_{w}(\vec{r},\vec{r}_{0};0)\nonumber \end{eqnarray}
 where $\alpha$ is a constant which we shall determine. The static
Green's function happens to be the potential which solves Poisson's
equation for a point charge between two grounded, conducting plates,\cite{MorseFeshbach}
and we can put it in closed form as follows:\begin{eqnarray}
G_{w}(\vec{r},\vec{r}_{0};0) & = & -\sum_{m=1}^{\infty}\frac{d}{m\pi}\chi_{m}(y)\chi_{m}(y_{0})e^{-\frac{m\pi}{d}\left|x-x_{0}\right|}\label{eq:gw1}\\
 & = & -\frac{1}{\pi}\sum_{m=1}^{\infty}\frac{d}{m}\times\frac{2}{d}\sin\left(\frac{m\pi y}{d}\right)\label{eq:gw2}\\
 &  & \hspace{.25in}\times\sin\left(\frac{m\pi y_{0}}{d}\right)e^{-\frac{m\pi}{d}\left|x-x_{0}\right|}\nonumber \\
 & = & \frac{1}{\pi}\sum_{m=1}^{\infty}\frac{1}{m}\left[\cos\left(\frac{m\pi(y+y_{0})}{d}\right)\right.\label{eq:gw3}\\
 &  & \hspace{.25in}\left.-\cos\left(\frac{m\pi(y-y_{0})}{d}\right)\right]e^{-\frac{m\pi}{d}\left|x-x_{0}\right|}\nonumber \\
 & = & \frac{1}{\pi}\textrm{Re }\sum_{m=1}^{\infty}\frac{1}{m}e^{-\frac{m\pi}{d}\left|x-x_{0}\right|}\label{eq:g24}\\
 &  & \hspace{.5in}\times\left[e^{\frac{im\pi(y+y_{0})}{d}}-e^{\frac{im\pi(y-y_{0})}{d}}\right]\nonumber \\
 & = & \frac{1}{\pi}\textrm{Re }\sum_{m=1}^{\infty}\left(\frac{Z_{+}^{m}}{m}-\frac{Z_{-}^{m}}{m}\right)\label{eq:gw5}\end{eqnarray}
 where\begin{equation}
Z_{\pm}\equiv e^{-\frac{\pi}{d}\left|x-x_{0}\right|}e^{\frac{i\pi(y\pm y_{0})}{d}}.\end{equation}
 Using the identities\begin{equation}
\sum_{m=1}^{\infty}\frac{Z^{m}}{m}=-\ln(1-Z)\end{equation}
 and\begin{equation}
{\rm Re}\ln Z=\ln|Z|\end{equation}
 we find\begin{eqnarray}
G_{w}(\vec{r},\vec{r}_{0};0) & = & \frac{1}{\pi}\ln\left|\frac{1-Z_{-}}{1-Z_{+}}\right|\label{eq:limit1}\\
 & = & \frac{1}{2\pi}\ln\left(\left|\frac{1-Z_{-}}{1-Z_{+}}\right|^{2}\right)\label{eq:limit2}\\
 & = & \frac{1}{2\pi}\ln\left|\frac{e^{\frac{\pi}{d}\left|x-x_{0}\right|}-e^{\frac{i\pi}{d}(y-y_{0})}}{e^{\frac{\pi}{d}\left|x-x_{0}\right|}-e^{\frac{i\pi}{d}(y+y_{0})}}\right|^{2}\label{eq:limit3}\\
 & = & \frac{1}{2\pi}\ln\left[\frac{\left(e^{\frac{\pi}{d}\left|x-x_{0}\right|}-e^{\frac{i\pi}{d}(y-y_{0})}\right)}{\left(e^{\frac{\pi}{d}\left|x-x_{0}\right|}-e^{\frac{i\pi}{d}(y+y_{0})}\right)}\right.\label{eq:limitform}\\
 &  & \hspace{.25in}\left.\times\frac{\left(e^{\frac{\pi}{d}\left|x-x_{0}\right|}-e^{-\frac{i\pi}{d}(y-y_{0})}\right)}{\left(e^{\frac{\pi}{d}\left|x-x_{0}\right|}-e^{-\frac{i\pi}{d}(y+y_{0})}\right)}\right]\nonumber \\
 & = & \frac{1}{2\pi}\ln\frac{\cos\left[\frac{\pi}{d}(y-y_{0})\right]-\cosh\left[\frac{\pi}{d}(x-x_{0})\right]}{\cos\left[\frac{\pi}{d}(y+y_{0})\right]-\cosh\left[\frac{\pi}{d}(x-x_{0})\right]}.\nonumber \end{eqnarray}

The singularity of the Green's function arises from a logarithmic
singularity in the Hankel term at the source. Near the scatterer,
the contribution from this term becomes\begin{eqnarray}
\lim_{r\rightarrow r_{0}}G_{0}(r,r_{0};k) & = & -\frac{i}{2}\lim_{\vec{r}\rightarrow\vec{r}_{0}}H_{0}(k\left|\vec{r}-\vec{r}_{0}\right|)\label{eq:b15}\\
 & = & -\frac{i}{2}+\frac{1}{\pi}\ln\left(\frac{k}{2}\left|\vec{r}-\vec{r}_{0}\right|\right)\label{eq:b16}\\
 &  & \hspace{.5in}+\frac{1}{\pi}\gamma\nonumber \\
 & = & -\frac{i}{2}+\frac{\gamma-\ln2}{\pi}\label{eq:b17}\\
 &  & \hspace{.5in}+\frac{1}{\pi}\ln\left(k|r-r_{0}|\right)\nonumber \end{eqnarray}
 where $\gamma$ is the Euler-Mascheroni constant, while from Eq.
(\ref{eq:limitform}), we find that the limiting behavior of the static
Green's function is (after some algebra)\begin{eqnarray}
\lim_{\vec{r}\rightarrow\vec{r}_{0}}G_{w}(\vec{r},\vec{r}_{0};0) & = & -\frac{1}{\pi}\ln\frac{2kd}{\pi}\sin\left(\frac{\pi y_{0}}{d}\right)\label{eq:limit5}\\
 &  & +\frac{1}{\pi}\ln\left(k|r-r_{0}|\right).\nonumber \end{eqnarray}
In order to cancel the logarithmic singularity in Green's function,
we thus require that\begin{equation}
\alpha=1\label{eq:gamma}\end{equation}
 so that\begin{eqnarray}
\lim_{\vec{r}\rightarrow\vec{r}_{0}}\left[G_{w}(\vec{r},\vec{r}_{0};0)-G_{0}(\vec{r},\vec{r}_{0};k)\right] & =\nonumber \\
 &  & \hspace{-1.5in}-\frac{1}{\pi}\ln\left[\frac{kd}{\pi}\sin\left(\frac{\pi y_{0}}{d}\right)\right]+\frac{i}{2}-\frac{\gamma}{\pi}\label{eq:limit6}\end{eqnarray}
Our final form for the Green's function, using (\ref{eq:quickconverge}-\ref{eq:gamma})
is\begin{eqnarray}
G_{w}(\vec{r},\vec{r}_{0};k) & = & \sum_{m=1}^{\infty}\chi_{m}(y)\chi_{m}(y_{0})\left(\frac{1}{ik_{x}^{(m)}}e^{ik_{x}^{(m)}|x-x_{0}|}\right.\nonumber \\
 &  & \hspace{.5in}\left.+\frac{d}{m\pi}e^{-\frac{m\pi}{d}|x-x_{0}|}\right)\label{eq:limit7}\\
 &  & +\frac{1}{2\pi}\ln\frac{\cos\left[\frac{\pi}{d}(y-y_{0})\right]-\cosh\left[\frac{\pi}{d}(x-x_{0})\right]}{\cos\left[\frac{\pi}{d}(y+y_{0})\right]-\cosh\left[\frac{\pi}{d}(x-x_{0})\right]}.\nonumber \end{eqnarray}
This expression is of further use in that it allows us to obtain another
form of the renormalization constant (\ref{eq:lambda}), which we
had shown to have the slowly convergent expansion\begin{equation}
G_{r}=-\frac{i}{2}\sum_{m=1}^{\infty}(-1)^{m}H_{0}(k|\vec{r}_{m}-\vec{r}_{0}|).\end{equation}
We can now express $G_{r}$ in an equivalent, but more rapidly convergent
expression, suitable for numerical purposes:\begin{eqnarray}
G_{r} & = & \lim_{\vec{r}\rightarrow\vec{r}_{0}}\left(G_{w}(\vec{r},\vec{r}_{0};k)-G_{0}(\vec{r},\vec{r}_{0};k)\right)\label{eq:limit8}\\
 & = & \lim_{\vec{r}\rightarrow\vec{r}_{0}}\left[\left(G_{w}(\vec{r},\vec{r}_{0};k)-G_{w}(\vec{r},\vec{r}_{0};0)\right)\right.\nonumber \\
 &  & \hspace{.25in}\left.+\left(G_{w}(r,r_{0};0)-G_{0}(r,r_{0};k)\right)\right]\label{eq:limit9}\\
 & = & \sum_{m=1}^{\infty}\left(\frac{1}{ik_{x}^{(m)}}+\frac{d}{m\pi}\right)\chi_{m}^{2}(y_{0})\nonumber \\
 &  & \hspace{.25in}-\frac{1}{\pi}\ln\left[\frac{kd}{\pi}\sin\left(\frac{\pi y_{0}}{d}\right)\right]+\frac{i}{2}-\frac{\gamma}{\pi}.\label{eq:kummerform}\end{eqnarray}

\end{document}